# CHAPTER 4

# SPIRAL GALAXIES

F. Combes, Observatoire de Paris, LERMA,
Collège de France, CNRS, PSL Univ., Sorbonne Univ., Paris, 75014, France

*Our vision of galaxies has changed significantly since the era of large galaxy surveys like the Sloan, which gave us extensive statistics with millions of galaxies. The Hubble sequence classification described in Chapter 1 still remains very widely used but has been enriched with broad categories based on color that indicate the recent formation of stars: the red sequence of passive galaxies, consisting solely of old stars, and the blue cloud of galaxies with active star formation. Chapter 3 focused on galaxies with a dominant spheroid, which are generally found on the red sequence. One of the key questions about the evolution of galaxies that remain to be answered is to understand how a galaxy can suddenly pass from one category to another.*

## 1    Introduction

This chapter is dedicated to spiral galaxies, also known as "late-type galaxies," which are dominated by their disc component. Some of them also have a spheroid component, or bulge, and the bulge/disc ratio increases along the Hubble sequence, from Sc to Sa. At this point, they then transition toward early-type galaxies dominated by a spheroid component, the majority of which are found on the red sequence. How do these two components form?

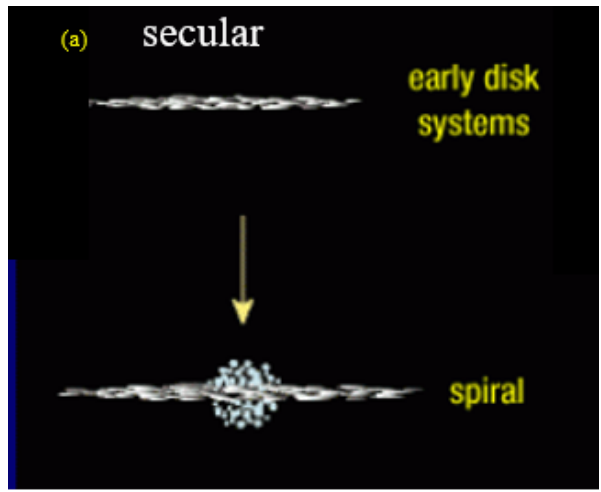

**Figure 1:** (a) Secular evolution of galaxies: when gas falls into a dark matter halo, the first systems to form are discs if stars begin to form only slowly, following gas dissipation. Internal evolution and the formation of a bar then form a pseudo-bulge by vertical resonance.

Galaxies begin by forming their dark matter component: the halos collapse under their own gravity, well before the recombination of the Universe 380,000 years after the Big Bang. Before this date, ordinary matter (or baryons) is ionized and interacts closely with photons, which stabilize it. Only the dark matter itself does not interact directly with the photons; it may collapse once matter dominates the radiation in terms of density. During recombination, the baryons decouple from the photons, possibly falling into the potential wells of the halos. These baryons primarily take the form of hydrogen gas, with 25% helium by mass. As it dissipates, the gas quickly forms rotating discs that conserve a large proportion of their initial angular momentum, acquired by the halos due to tidal torques. Local gravitational instability causes the gas to form stars, which also spread throughout the disc. Thus, the first baryonic components naturally form as discs, as shown in Figure 1.

Figure 1(b) shows the hierarchical scenario in which spheroid components may form. When two disc galaxies with randomly oriented angular momenta merge, the final angular momentum is logically expected to be smaller, as a vector sum of randomly oriented momenta. Multiple minor mergers can create an increasingly massive bulge, whereas major mergers of spiral galaxies (i.e. mergers with a mass ratio close to 1) will form elliptical galaxies that are dominated by their spheroid. However, another scenario is possible, depending on the speed and efficiency of star formation during the gas collapse in the halos. If the star formation occurs more quickly than the collapse time, there is a so-called "monolithic" scenario, where the spheroid component forms first, and the spheroidal galaxy then potentially forms a disc later by the external accretion of gas originating from cosmic filaments (cf. Figure 1(b), left).

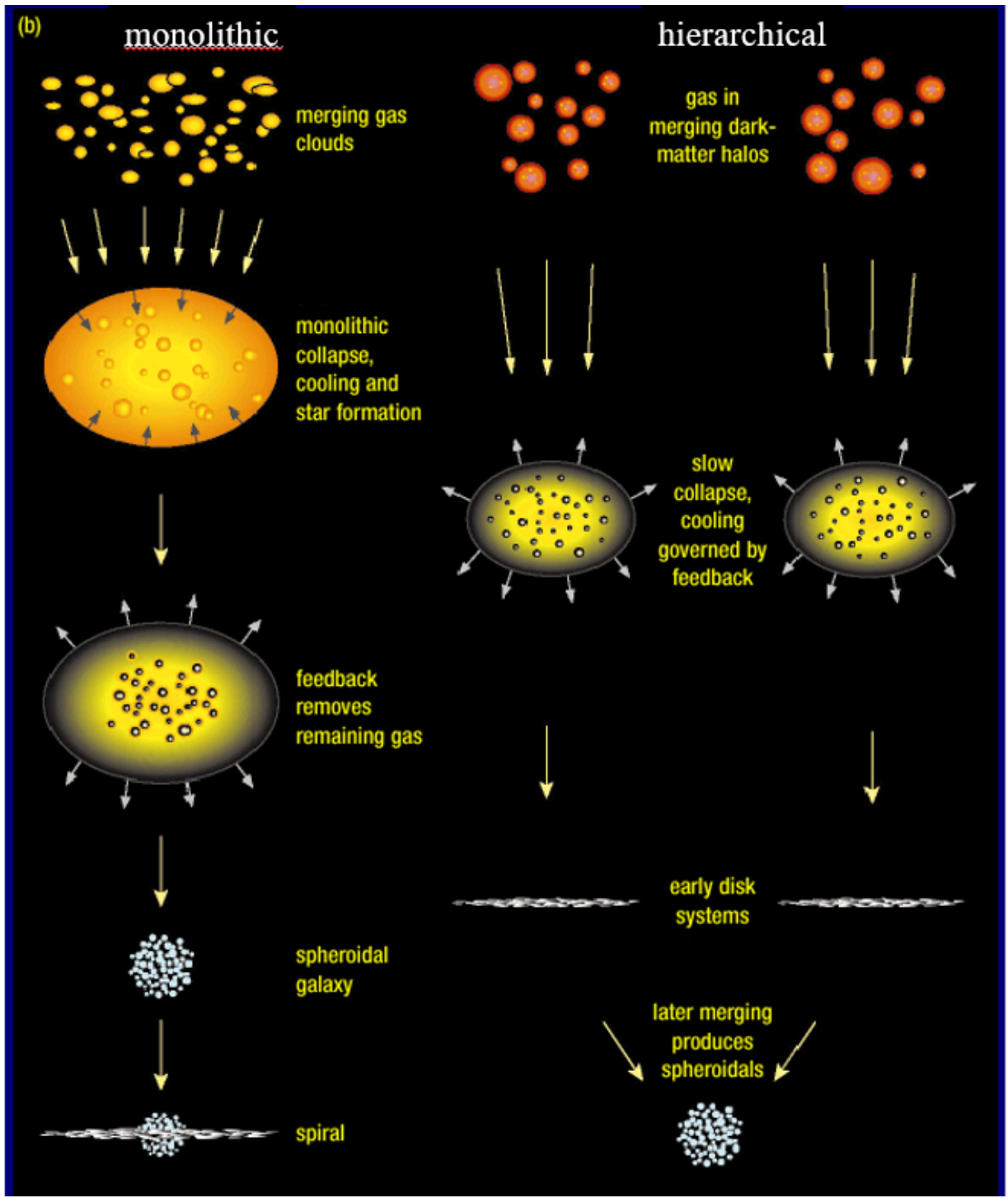

**Figure 1:** (b) Two other scenarios that may create spheroids. On the right, the hierarchical scenario: once both discs have formed, the interaction and merger of the two spiral galaxies may create an elliptical galaxy, or at least a galaxy dominated by its spheroid component. On the left, if the star formation occurs more quickly than the dissipation of gas, the stars form with a spheroidal geometry straight away. This is the monolithic collapse scenario.

Monolithic collapse was in fact one of the first scenarios considered to explain the formation of our own galaxy, the Milky Way, by Eggen, Lynden-Bell, and Sandage (1962). The idea came from observing the galaxy's stellar halo, which consists of very old stars that are poor in heavy elements. These metals (like any elements that did not form in the Big Bang, such as carbon, oxygen, and iron) are ejected by the earliest stars, enriching the interstellar medium where the stars of second and later generations then form. Metallicity is therefore an indicator of age, or number of astration cycles. Furthermore, Eggen, Lynden-Bell, and Sandage (1962) observed that, for stars in the halo, there is a correlation between the metallicity, indicated by [Fe/H], and the eccentricity. They interpreted this correlation as a change in the dynamics as the gas slowly collapses into a disc. The stars that form the earliest have a three-dimensional geometry and are poor in metals. Gradually, their orbits become less radial, tending toward circular orbits in the equatorial plane of the galaxy. In this scenario, the gravitational potential only evolves slowly, which allows us to consider the eccentricity and the angular momentum $L_z$ perpendicular to the disc as invariants. However, Searle & Zinn (1978) were not convinced by these hypotheses, objecting that the globular clusters of the stellar halo show large metallicity discrepancies that do not correlate with the distance to the center of the galaxy. Instead, they favored the explanation of a merger of small protogalaxies with a rapidly varying gravitational potential. Today, this scenario has prevailed, and it is widely accepted that many of the stars in the halo of the Milky Way originate from minor mergers. Numerical simulations have shown that stellar streams can remain visible even 10 billion years after the destruction of a small satellite galaxy (e.g. Helmi & White 1999).

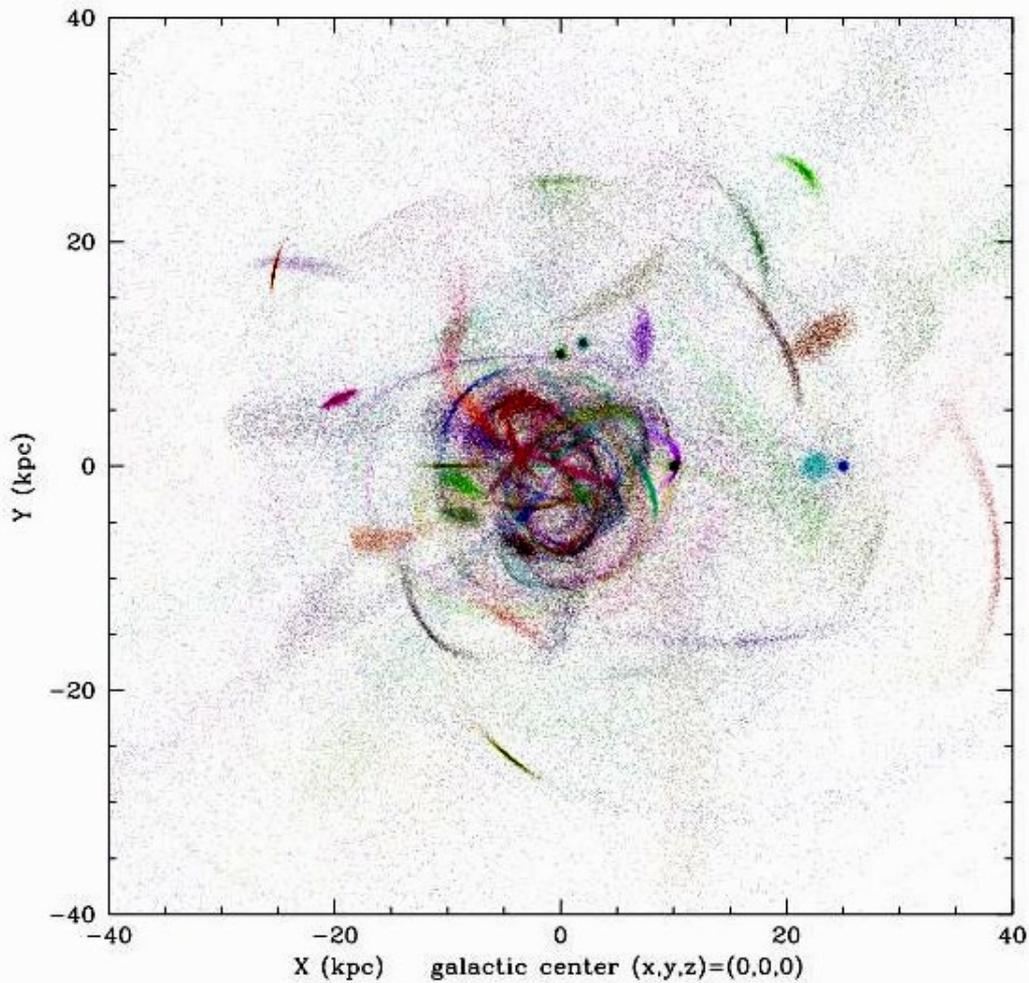

**Figure 2:** The stellar halos of the Milky Way and Andromeda reveal evidence of their formation by accretion and destruction of satellite dwarf galaxies and globular clusters in the form of tidal loops and streams. They are entirely comparable to this simulation, which shows the accretion of 50 dwarf galaxies over a period of 10 billion years. Credit: P. Harding, 2001, PhDT.

The original "monolithic" scenario has been abandoned, but another more elaborate version is still invoked to explain the extreme formation speed of some massive elliptical galaxies observed at high redshifts. It is not yet clear what proportion of galaxies form in this way (monolithic scenario) as opposed to by mergers of spiral galaxies (hierarchical scenario), which could unfold very quickly in overdense environments, ancestors of the dense galaxy clusters of today. However, it has been clearly established that spiral galaxies completely dominated by their disc component, either with a very small bulge or no bulge at all, are the field galaxies most frequently observed today, and that they could only have formed according to the secular scenario of Figure 1(a).

Today, observations from large surveys of local galaxies, together with deep cartographies sampling distant galaxies, have revealed to us the history of the cosmic star formation (e.g. Madau & Disckinson 2014). Starting from the Big Bang, the star formation density increases regularly for around 4 billion years, up to a peak at $z=2$, then decreases regularly by a factor of 20 since this peak. Today, most galaxies are passive, while they had very active star formation for the first half of the history of the Universe. This also reflects the bimodality between the red sequence and the blue cloud highlighted below. Clearly, active star formation suddenly stopped in many galaxies, and it is essential for us to understand the mechanism of this change.

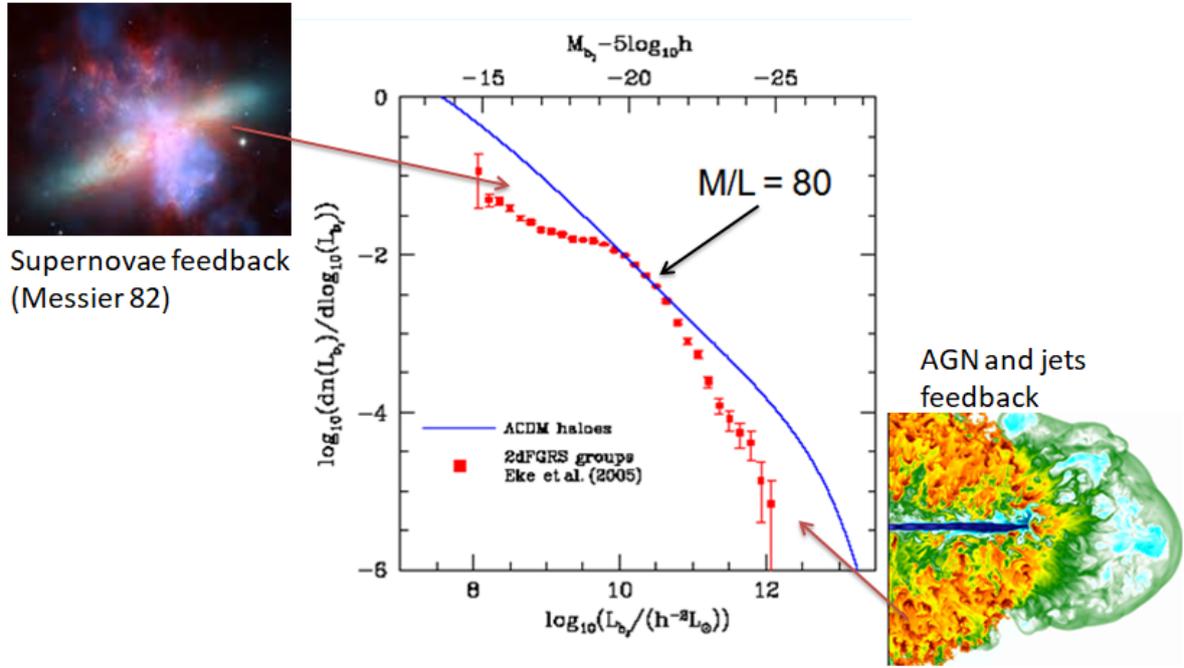

**Figure 3:** Distribution of galaxies by luminosity (log-log diagram): the red points with error bars are the observations (Eke et al 2006), and the blue curve shows the predictions of the ΛCDM standard cosmological model. The two distributions are calibrated at their point of intersection, with a mass-to-luminosity ratio of M/L= 80. The model predicts too many small galaxies, which can be corrected with supernovae feedback (left), and too many massive galaxies, which could be prevented by feedback from active galaxy nuclei (right).

This discussion is informed by another observation: the luminosity function of galaxies, which is shown in Figure 3 (cf. Jenkins et al 2001, Baugh 2006). The number of galaxies with low stellar mass (up to $M_* = 3·10^{10} M_\odot$) is smaller than the number predicted by numerical simulations performed within the framework of the standard cosmological model, ΛCDM. The number of galaxies that are more massive than this limit is also lower than expected. The central point is used for calibration and corresponds to a mass-to-luminosity ratio of 80, much higher than expected from the universal ratio of total matter (dark+ordinary) to baryons, which is equal to 6. This calibration reveals that most baryons have exited the galaxies and are now located in intergalactic space, more precisely in cosmic filaments. These observations all tend to show that galaxy formation is accompanied by violent events that are capable of ejecting the majority of the visible matter out of the halos. For low-mass galaxies, feedback from star formation, i.e. supernova explosions and stellar wind, can eject the surrounding gas (Dekel & Silk 1986). But this mechanism is not sufficient for massive galaxies, as the gas does not reach escape velocity. The energy deployed by the active nuclei around supermassive black holes at the center of each galaxy might give us the answer (Croton et al 2006).
These two processes must certainly play an important role in the formation of galaxies. However, empirically, the two observational parameters that seem to control the star formation activity of galaxies are the total mass and the environment, namely whether the galaxy belongs to a group or cluster or has an average galaxy density in its immediate neighborhood (Peng et al 2010). Is the total mass of baryons that accumulates at the center of a galaxy capable of stabilizing the disc and preventing the formation of stars? Or does the environment rip the gas away from the galaxy, or perhaps prevent cosmic filaments from resupplying it?

We will examine these various mechanisms that control the star formation, then turn our attention to the spiral arm structure of the discs, as well as the formation of bars, which are key drivers of the evolution of these discs and the assembly of their mass. We will also see how the environment can play a crucial role.

## 2  Blue and red galaxies: quenching star formation

### 2.1  Definition of bimodality

If we plot the hundreds of thousands of local galaxies of the Sloan survey on a color-magnitude diagram, we find that they are not distributed homogeneously. Instead, they occupy certain clearly distinct regions. This is called bimodality (Baldry et al 2004). In the stellar mass-color diagram of Figure 4, we can clearly see a red sequence, which encompasses the most massive galaxies, and a blue cloud, in the region with lower masses. This diagram can be decomposed by galaxy morphology according to whether the spheroid component dominates the disc component and whether the disc has spiral arms. The red sequence clearly corresponds to morphological types that are dominated by their spheroid, namely the so-called early-type galaxies (ETGs, see Chapter 3). The blue cloud corresponds to spiral galaxies and late-type galaxies, although it is worth noting that some of the most massive spirals edge back toward the red sequence. These galaxies have the most massive bulges, and their star formation activity has stopped. For these galaxies, the divide between the two sequences, known as the green valley (Schawinski et al 2014), is less pronounced than for all galaxies combined.

The bimodality of the color-magnitude diagrams shows the key importance of the star formation rate and its history in the evolution of a galaxy, as well as the importance of the quantity of gas and dust, which is linked to the metallicity. It sheds light on two mechanisms of galaxy formation, with a separation corresponding to the critical stellar mass of $3 \cdot 10^{10} \, M_\odot$.

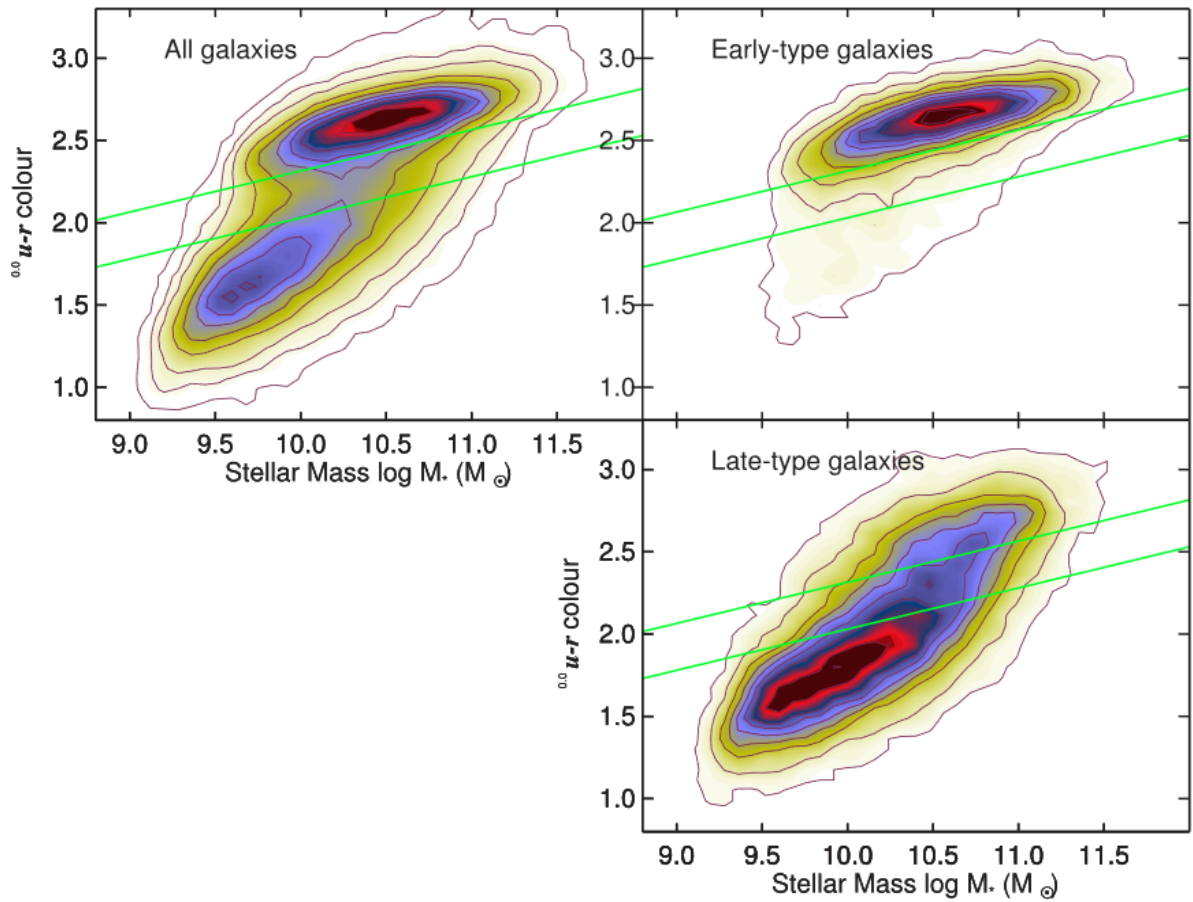

**Figure 4:** Color (u-r)-stellar mass diagram for all galaxies combined (top left), and for early-type and late-type galaxies separately. The green lines show the position of the green valley, which separates the red sequence above it from the blue cloud below it. Reproduced from Schawinski et al (2014).

Large-scale statistics about the properties of galaxies have been collected over the past few years, not just from large surveys like Sloan and DECaLS (Dark Energy Camera Legacy Survey), but also thanks to a voluntary classification by the keen amateur astronomers who participated in the great "Galaxy Zoo" adventure (Lintott et al 2008). Each person classifies a few hundred galaxies. Ultimately, each galaxy receives around a dozen responses, which allows any mistakes to be eliminated. On a small test sample, the classification produced by these amateurs coincided with that of experts. A million galaxies have been added to the classification since July 2017 in version 4 of the Zoo. For more distant galaxies, deep surveys by the Hubble Space Telescope (HST) are used, revealing a major evolution as the redshift increases. Distant galaxies become increasingly irregular, with clumps, and the Hubble sequence is only applicable to a small minority of massive galaxies in the first half of the history of the Universe. Today, automatic classifications are becoming more and more sophisticated. Artificial intelligence will represent a very powerful tool as soon as deep learning programs are ready (e.g. Dominguez Sanchez et al 2018). Learning is often performed on cosmological simulations.

As shown in Figure 4, the colors do not fully match the Hubble sequence. We can define various parameters, such as the surface density µ (in magnitude per second squared), or the Sersic index n, i.e. the slope of the radial distribution of light in the galaxy. The index n is defined by $I(R) = I_0 \exp(-kR^{1/n})$, where $I(R)$ is the light surface density averaged along the azimuth, and $I_0$ is its value at R=0. Figure 5 shows the correspondence between a simplified Hubble sequence (elliptical, early-type spiral, and late-type spiral), the color (u-r), the Sersic index n, and the surface brightness µ. Clearly, galaxies are distributed bimodally by color, but also by the index n and the surface brightness µ. The red and passive galaxies are for the most part elliptical galaxies, which are traditionally characterized by the De Vaucouleurs light profile with n=4, and which are more compact (low µ). Late-type spiral galaxies are blue, and their discs satisfy the relation described by Freeman in 1970 ($\mu_0$ = 21 mag/arcsec$^2$). The separation is less pronounced for early-type spiral galaxies, which are found at the transition.

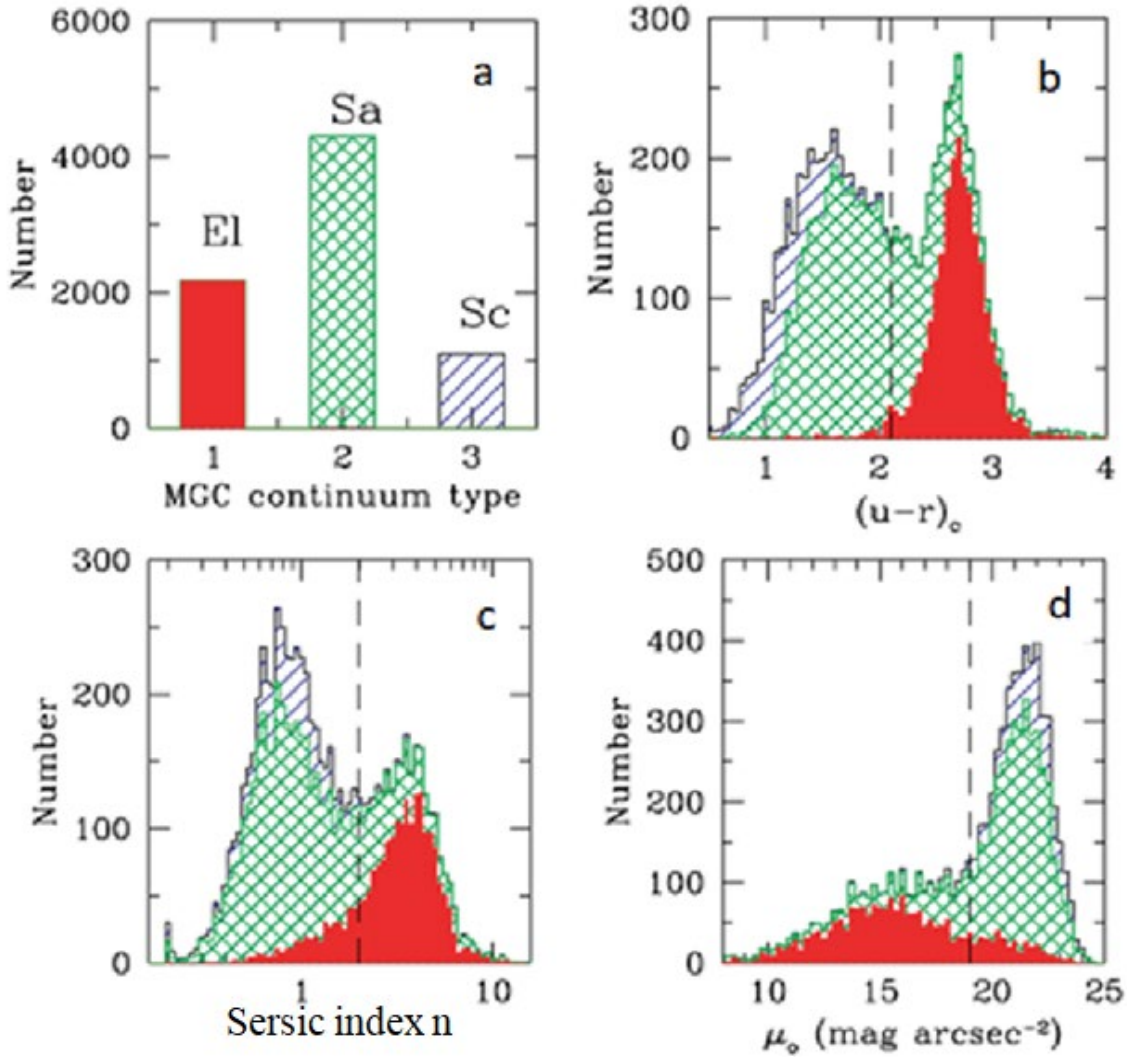

**Figure 5:** Histograms of the number of galaxies by morphological Hubble type (a), color u-r (b), Sersic index n (c), and central surface brightness $\mu_0$ for galaxies in the MGC catalog. Reproduced from Driver et al. (2006).

Observe that, for local galaxies, the most active galaxies in terms of star formation are the least massive, namely the late-type spiral galaxies and the irregular galaxies. This trend is part of a more general phenomenon, known as "downsizing" or hierarchical inversion (Cowie et al 1996): the luminosity of the most active galaxies continuously decreases over time. Indeed, according to the standard cosmological model, which is hierarchical, we expect the earliest objects that form to be the smallest; these objects then merge to form larger objects, and, today, we would expect to observe the formation of the most massive galaxies. But this is not the case at all. However, this paradox is not quite so paradoxical if we consider that the hierarchy only concerns the dark matter halos. At the beginning of the Universe, the least massive halos collapse first. By merging, they then form larger and larger halos. Overdensities correspond to groups that collapse later, and, similarly, galaxy clusters virialize at around $z=1-2$. Today, most clusters continue to grow through mergers, and we frequently see multiple distinct subclusters form into a rich cluster, like the Coma Cluster. Baryons, on the other hand, have another fate. At the beginning of the Universe, gas falls into the halos, forming dwarf galaxies and, gradually, stars. The gas consumed in these stars is partially replaced by accretion from cosmic filaments, but the density of these filaments decreases as the expansion progresses, and the fraction of gas in the galaxies also decreases over time. As a result, galaxy mergers involve old stars more and more frequently, without any gas or a new generation of stars. The most massive galaxies are therefore found in the red sequence and are passive, even if they were recently involved in a merger.

## 2.2 The parameters that determine the red sequence

The fraction of "red" galaxies with inactive star formation depends on several parameters. We can already see on Figure 4 that the red sequence encompasses the most massive galaxies. It turns out that there are two parameters that are strongly correlated with the color – not just the mass, but also the richness of the environment, namely the surface density $\Sigma$ after projecting the neighboring galaxies. Figure 6 clearly shows these two effects. The fraction of red galaxies increases as a function of the stellar mass for each curve of any given color, which corresponds to a fixed surface density $\Sigma$ of galaxies per Mpc$^2$. From these observations, we can now deduce the physical phenomena that cause quenching of star formation within a galaxy.

The angular momentum is an essential parameter in the fate of a galaxy. It is possible to measure the rotation curves of disc galaxies, as well as the rotation curves of spheroid galaxies, even though the latter rotate very little, since they are in equilibrium due to their anisotropic velocity dispersion. Given the rotation velocity at each radius, we can calculate the total angular momentum, and, more significantly, the specific momentum j per unit mass. Fall & Romanowsky (2013) showed that the specific momentum j can be used as another parameter to classify galaxies: every galaxy in the Hubble sequence is aligned along parallels with slope 2/3 in the logj-logM∗ diagram when only baryonic quantities are considered. The spiral galaxies of latest type are found at the top of the diagram. Many other versions of this diagram and classification have been published (e.g. Sweet et al 2018). Some incorporate a third parameter, the bulge-to-disc ratio, which allows classical bulges resulting from mergers to be distinguished from pseudo-bulges, which result from resonance with a bar in the secular scenario.

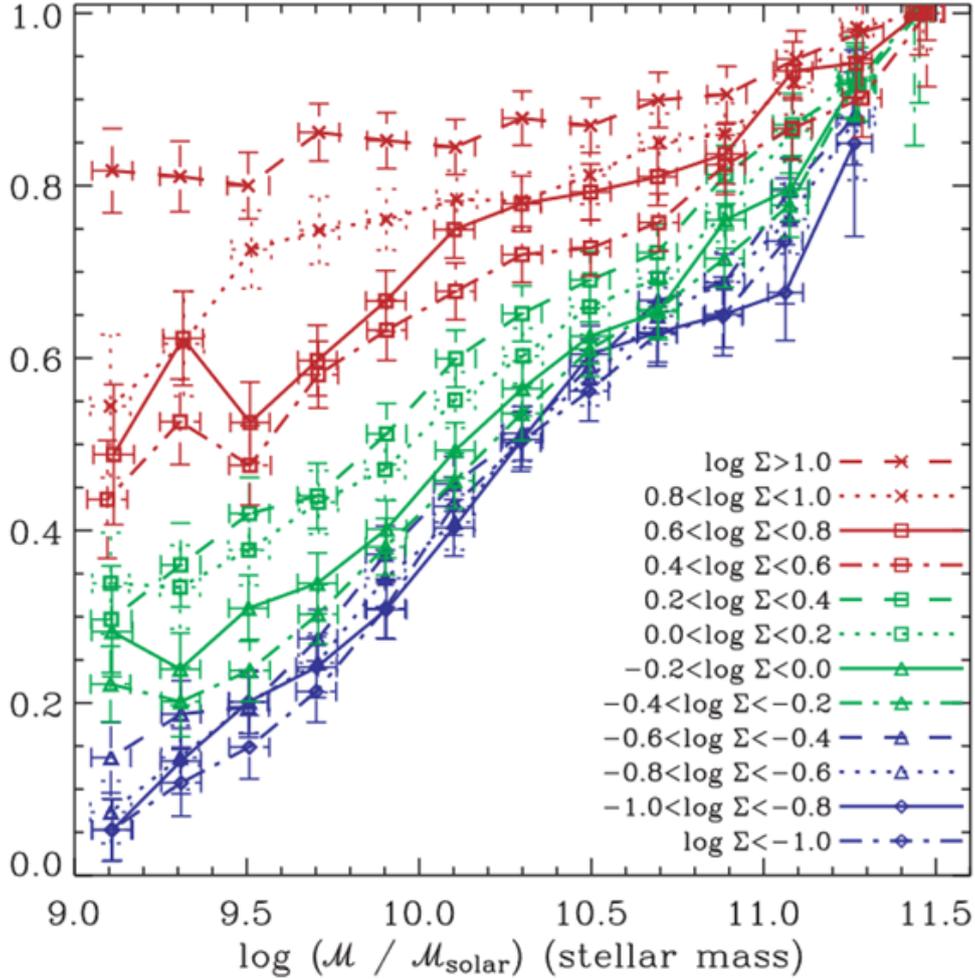

**Figure 6:** Fraction of red galaxies as a function of the stellar mass M, ranging from $10^9$ M$_\odot$ to $4 \cdot 10^{11}$ M$_\odot$, for several galaxy surface density values Σ ranging from 0.1 to 10 gal Mpc$^{-2}$. Reproduced from Baldry et al. (2006).

Of course, these parameters are not the only possible choices: the color of a galaxy is also a function of its gas fraction, the quantity of dust, the age of its stars, the metallicity, and the morphological type. Note that the star formation history of a galaxy depends on its surface density. As shown in Figure 7, galaxies are not distributed randomly on the diagram of central surface density versus stellar mass. Massive galaxies have a high surface density, and spiral galaxies satisfy the relation proposed by Freeman (1970). They are all HSB (High Surface Brightness). By contrast, dwarf galaxies, which have a low stellar mass, are mostly LSB (Low Surface Brightness) objects. Such objects are generally rich in atomic and possibly molecular gas. They should be capable of forming stars, but their efficiency is low. The massive (HSB) galaxies are more compact, with an older stellar population and a redder color. It is difficult to say which is cause and which is effect. If initial conditions such as angular momentum and high rotation prevent mass from concentrating, this may explain why the galaxy remains at a dwarf stage, with low surface brightness. The transition between the two categories occurs once again at the critical stellar mass of M$_*$=$3 \cdot 10^{10}$ M$_\odot$, or a surface density of $3 \cdot 10^8$ M$_\odot$/kpc$^2$. Ultimately, the star formation history may depend more on the surface density than the mass.

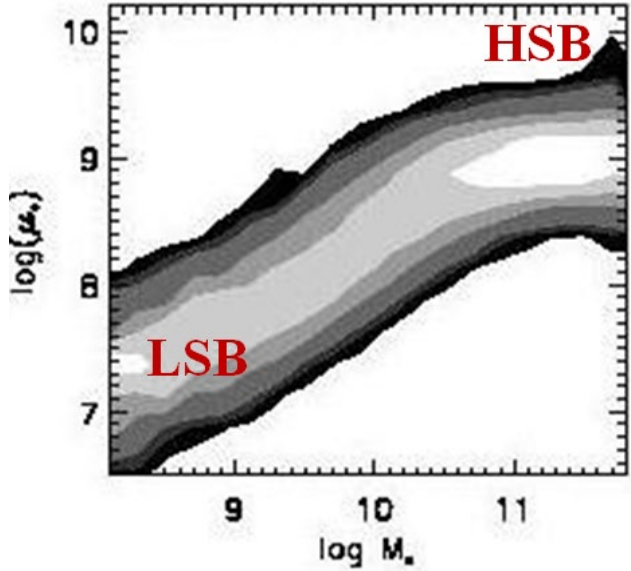

**Figure 7:** Surface density at galaxy centers as a function of the stellar mass M*. For massive galaxies with high surface brightness (HSB), Freeman's law (1970) is satisfied with a constant value for all discs. There is also a category of low surface brightness (LSB) galaxies, which are mostly dwarf galaxies. Diagram generated with 122,000 galaxies. Reproduced from Kauffmann et al (2003).

The transition around the critical mass of $M_*=3·10^{10}$ $M_\odot$ could reflect the importance of feedback from star formation. For shallow potential wells, the velocity of the gas ejected by supernovae, of the order of 100 km/s, is high enough to impart escape velocity to the dragged gas, and the galaxy loses its gas. By contrast, more massive galaxies retain their gas, and star formation accumulates.

## 2.3  Mechanisms for quenching star formation

Many mechanisms that might abruptly stop the formation of stars have been suggested. The occupied regions in the color-stellar mass diagrams of galaxies (Figure 4) suggest that this stopping does indeed occur suddenly. The number of galaxies between the red sequence and the blue cloud is low (green valley) and reflects the short time spent in the transition zone.
One of the first such mechanisms is suppression of the gas supply. Galaxies form in cosmic filaments consisting of dark matter and gas, and they continue to be supplied by matter that falls from these filaments over the course of their lifetime. If the gravitational potential of the galaxy is not too deep, numerical simulations show that the gas can supply the galaxy while staying relatively cold, which means that it remains available for star formation. By contrast, if the potential well exceeds a certain mass and depth, the gas experiences a shock wave that brutally heats it. Being diffuse and hot, it is no longer sufficiently dense to cool, and remains at the equilibrium of the virial theorem in the dark matter halo. Hot gas can only become dense enough to cool down again in dense clusters. In massive galaxies and groups, the gas remains very hot and is therefore not available for star formation. The critical transition mass is thought to be of the order of $M_h=5·10^{10}$ $M_\odot$ for the halo, or $M_*=3·10^{10}$ $M_\odot$ for the stellar mass (Dekel & Birnboim 2006).
This mechanism predicts that the separation between the red sequence and blue cloud in the color-stellar mass diagram will only occur at low redshifts, when the potential wells are deep enough to generate shocks. The red sequence should therefore only appear after z=1. However, this is not consistent with observations. Even at large redshifts, there are elliptical or early-type galaxies that are already red, no longer forming any stars. Consequently, there must be other mechanisms. Galaxies with active star formation define the so-called main sequence, since their formation rate (SFR) is nearly proportional to their stellar mass (cf. Figure 8). There are indeed bursts of star formation during interactions or mergers of galaxies, but 90% of cosmic star formation occurs on the main sequence.

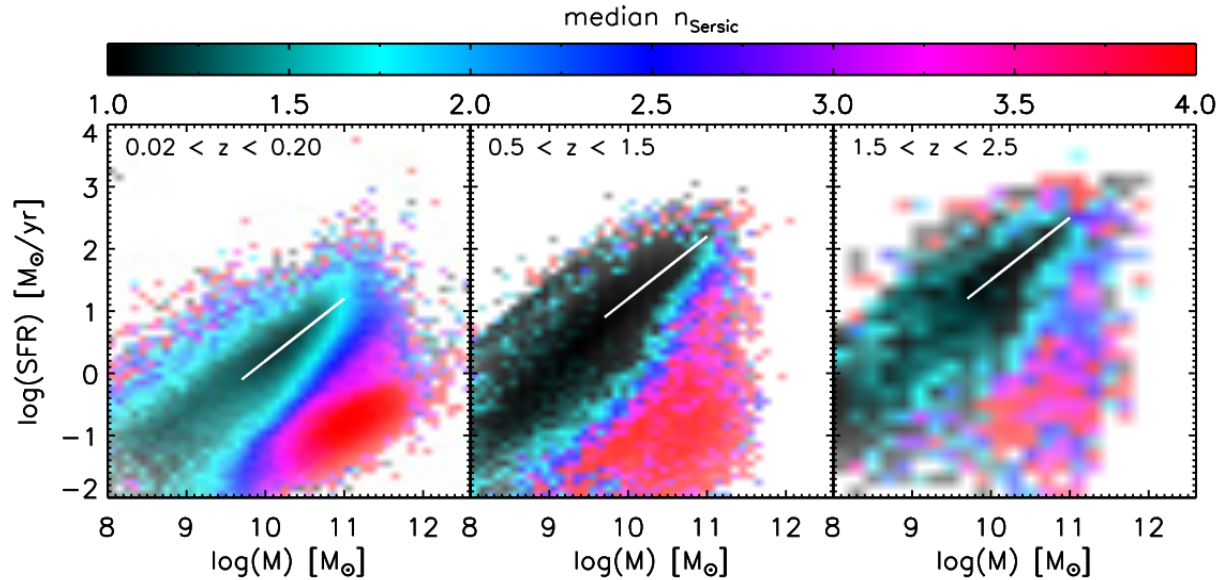

**Figure 8:** Distribution of galaxies in the diagram of star formation rate (SFR) and stellar mass (M∗), at three redshift ranges: on the left, the local galaxies of the Sloan survey, followed by 0.5 < z < 1.5 and 1.5 < z < 2.5, both from HST deep fields. The color indicates the Sersic index n of the radial light distribution. Blue corresponds to discs (n~1-2), and red corresponds to galaxies dominated by spheroid components (n>3). The blue galaxies are located on the main sequence, where SFR is proportional to M∗. The main sequence is indicated by a white line of slope 1 and zero point that increases with the redshift. At every epoch, the galaxies that form stars are discs, whereas the passive, red galaxies are spheroids. The less common systems above the main sequence form large numbers of stars and are primarily mergers or interactions of galaxies, with a spheroid profile. Reproduced from Wuyts et al (2011).

Observe that, in the color-stellar mass diagram (Figure 4), there are also low-mass galaxies on the red sequence. These galaxies must have ceased activity due to their interaction with a more massive companion; they are satellite galaxies (as opposed to the central galaxies).
Suppressing the cold gas supply of galaxies in very massive halos is a permanent but relatively slow way to stop star formation. Indeed, the central galaxy retains its cold gas and will continue to form stars until this gas is fully exhausted. This can take of the order of 2 billion years, the average time required to consume all the gas of a spiral galaxy at its current star formation rate (Bigiel 2008). We say that they have a depletion time of 2 Gyr. There are other, much quicker ways to stop star formation: for example, stripping of gas by ram pressure when a spiral enters a galaxy-rich cluster at high velocity, or during a violent collision between two galaxies. In these scenarios, the characteristic time is of the order of 100 million years.

Another mechanism that occurs over long time scales is stabilization of the gas in the disc by forming a very massive central spheroid (Martig et al 2009). This mechanism does not require the cold gas to be consumed, nor its supply to be discontinued. Even though cold gas is present in the disc, it will not form stars if the mass of the bulge is too high, provided that the gas's turbulence remains high, for example after a merger that created or increased the mass of the bulge. This mechanism is called "morphological quenching" because it only depends on the morphology. It applies to elliptical galaxies, and more generally to early-type galaxies, or in other words precisely the galaxies found in the red sequence.

In the simulation of Figure 9, a galaxy is followed in terms of cosmological conditions for a period equal to the Hubble time: accretion of gas from cosmic filaments, minor or major mergers. In an initial phase (t=3.5 Gyr), the galaxy is very rich in gas and unstable, which forms clumps, which is consistent with observations of galaxies at z>2. Then, after a major merger (t=10 Gyr), the galaxy experiences an inactive phase. The gas is stabilized by a massive bulge, as well as by the velocity dispersion (measured by the Toomre parameter Q). Eventually, the galaxy returns to a state with a colder gaseous disc, in particular by accretion. It experiences instabilities and begins to reform stars, like on the main sequence (t=13.7 Gyr, or z=0).

The Toomre parameter Q is an indicator of disc stability. It is equal to the ratio of the velocity dispersion $\sigma$ and the critical dispersion $\sigma_{crit}$ (Q= $\sigma/\sigma_{crit}$). The critical dispersion is defined in terms of the surface density $\Sigma$ of the disc and the epicyclic frequency $\kappa$, which is a function of the rotation velocity $\Omega$=V/R. Its value is $\sigma_{crit}$= 3.36 G$\Sigma$ /$\kappa$. The more the disc self-gravitates, the more it can be expected to be stabilized by the velocity dispersion. Conversely, a massive bulge causes $\kappa$ to increase, as well as Q if $\sigma$ and $\Sigma$ remain constant, which creates the stabilizing effect.

Finally, there are internal feedback processes, caused by either star formation or the activity of the central black hole (supernova feedback and AGN feedback). These processes can rapidly eject the gas, or heat it and make it turbulent, which stops star formation. These processes are rarely permanent. In massive galaxies, the gas does not reach escape velocity and falls back into the galaxy. As a result, the effect mostly just moderates the star formation, making the gas unavailable for a certain period of time. In dwarf galaxies, where the ejection velocity is greater than the escape velocity, the effect is temporarily stronger, but the galaxy may still accrete gas from cosmic filaments at a later point.

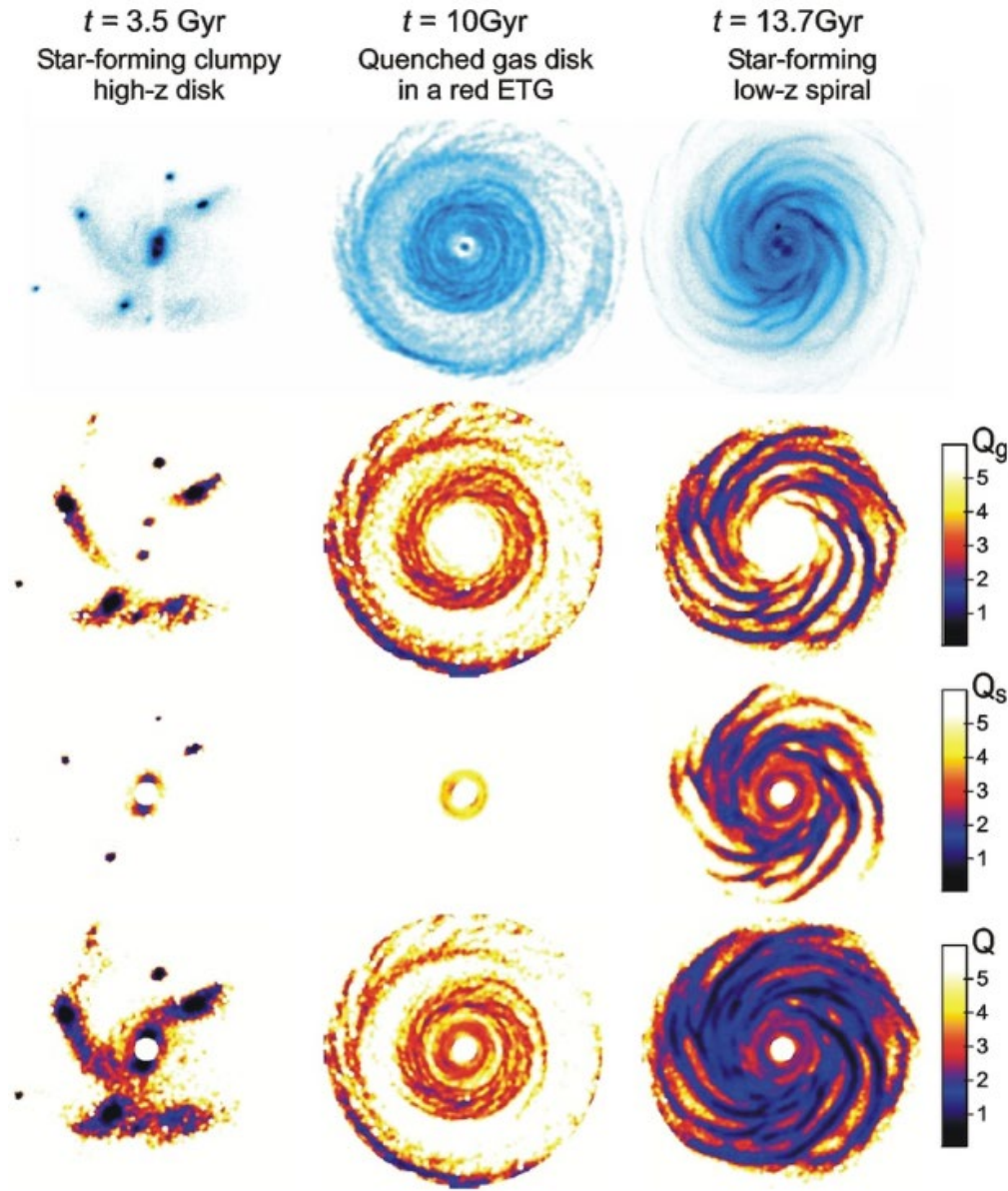

**Figure 9:** Gas distribution in a simulation that follows the life of a galaxy for a period equal to the Hubble time. The first row shows the gas surface density (in blue). The three other rows show the values of the Toomre parameter Q for the gas (Qg), for the stars (Qs), and for a combination of both (Q). The three columns correspond to three phases of the galaxy: a clumpy morphology, an inactive phase where the Q parameter can be seen to remain constantly high, and an active phase on the main sequence. From Martig et al (2009).

We can summarize the various mechanisms for stopping star formation by classifying them according to their speed: two categories of processes that unfold over large time scales and two categories of rapid processes:
(1) Processes that cut the cold gas supply of galaxies. The quenching time scales are larger (2-4 Gyr) because the galaxy continues to consume any gas that was already present. This may involve halos that are sufficiently massive at z<1 to heat gas falling into the galaxy by violent shocks. Effects from the environment (groups and clusters) can also cause the cold gas of filaments to completely disappear. The galaxy is suffocated, and the effect is permanent.

(2) Processes that stabilize the gas: these processes are slow, but not permanent. A massive and concentrated bulge forms from minor or major mergers and stabilizes the disc. This is a morphological process.

(3) Processes that eject the gas present in the galaxy. These processes are rapid (<100 Myr) but not permanent. This category includes feedback from supernovae and AGNs (winds or radio jets).

(4) Processes that heat the gas, rapid but very transitory. Turbulence may be generated by interactions between galaxies or feedback from supernovae. The gas dissipates its energy, and star formation resumes.

## 3 Spiral galaxies: density waves or not?

### 3.1 The winding problem

The morphology of disc galaxies is primarily distinguished by the existence of spiral arms. For a long time, spiral arms were a mystery, since they are usually coherent from one end of the galaxy to the other, winding around only less than once, like Messier 51 (cf. Figure 10).

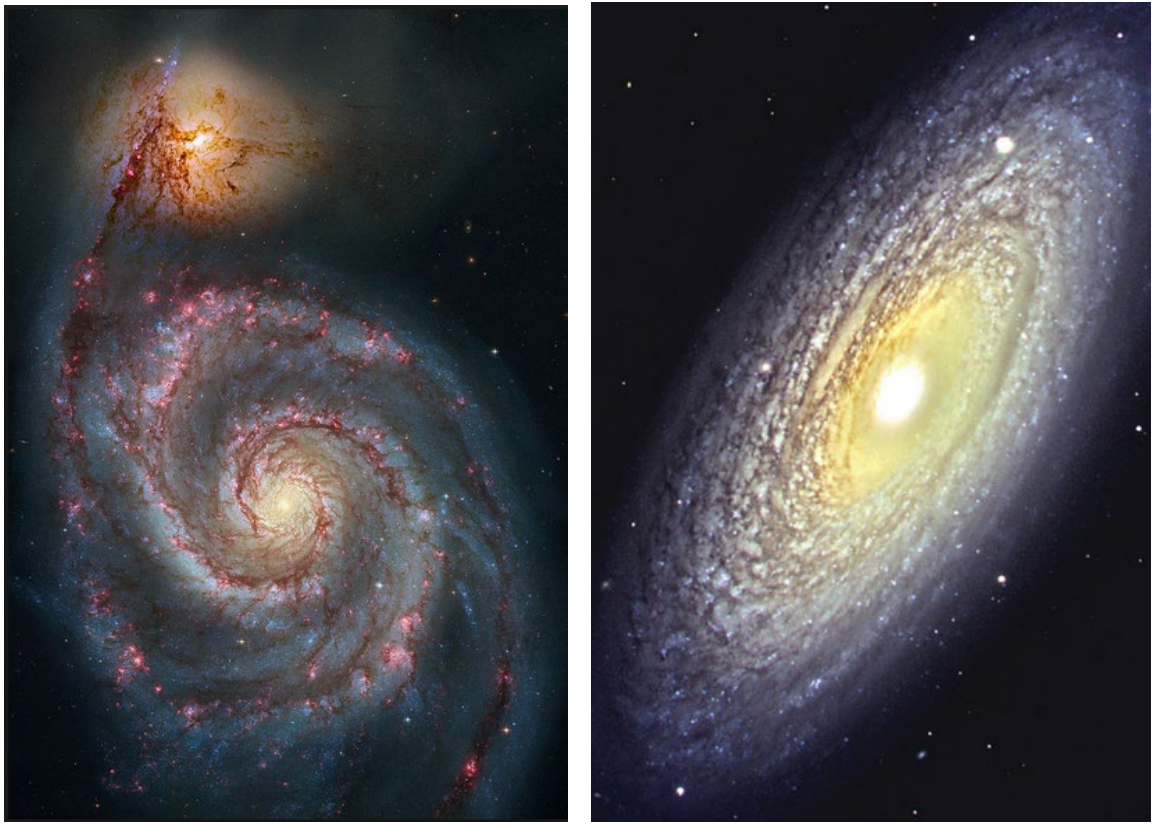

**Figure 10:** Two examples of spiral galaxies: on the left, Messier 51, which has two arms, winding just over 180°, very contrasted. This configuration is described as "grand design." On the right, NGC2841, a flocculent spiral galaxy with just spiral arms fragments, very tightly wound, but lacking coherence from one end of the galaxy to the other. This configuration is called a "stochastic spiral." It can be explained solely in terms of contagious star formation and differential rotation (Gerola & Seiden 1978). From HST (credit NASA/ESA).

Disc galaxies rotate around their center differentially. In other words, while the center completes one revolution in ~10 million years, the edges require a billion years to complete their rotation. There can be a factor of 100 between the rotation periods of the center and the edges. If the spiral arms of M51 were physical in nature, they should wind around at least 100 times, causing them to dissolve. How can this winding problem be solved?

The problem was solved in the 1960s to 1980s by the theory of density waves. The coherent spiral arms are not physical arms, but waves. All matter, whether stars or gas, passes through the arms during each revolution, staying proportionally longer within the spiral arms than in between them, resulting in the impression of an accumulation of matter. Bertil Lindblad was the first to suggest density waves after observing that the precession rate of the almost elliptical orbits of stars, $\Omega-\kappa/2$, remains almost constant as the radius increases, unlike the rotational angular momentum $\Omega$, which varies like $1/R$ for the flat rotation curves of galaxies (cf. Figure 11).

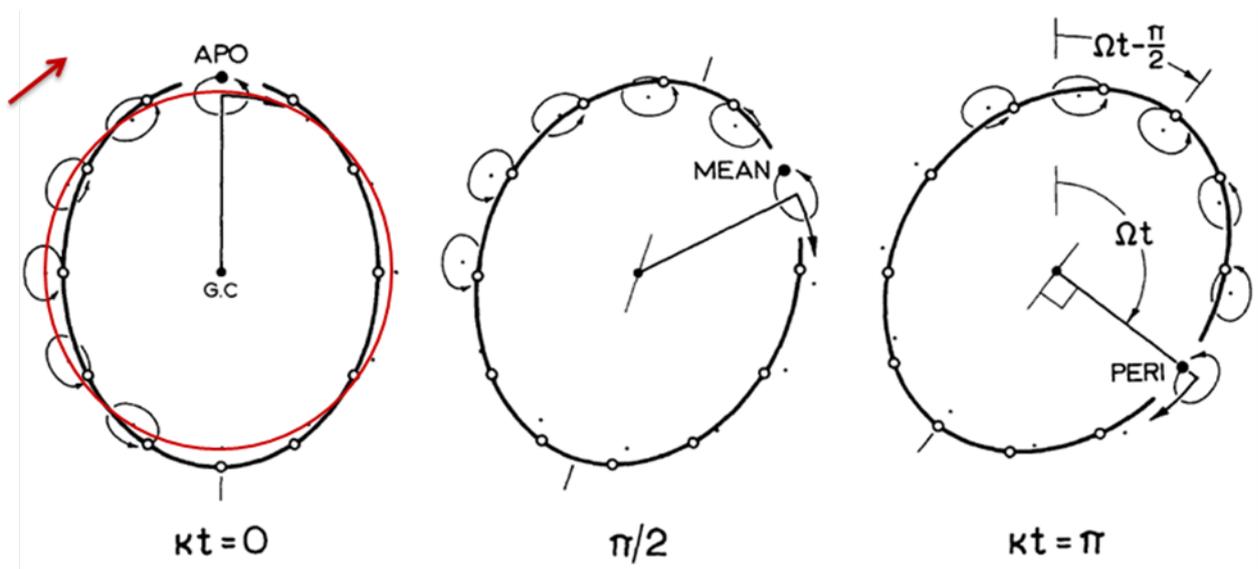

**Figure 11:** A typical stellar orbit in the epicyclic approximation. At order zero, the star completes a circle (in red). At the first order of perturbation, a small oscillation is added, with epicyclic frequency $\kappa$. At any given time, the star is represented by a small empty circle that completes an epicycle around the guide center, represented by a point (still on the red circle, left). The position of the star at three epochs is marked with a solid circle ($\kappa t=0$, apocenter, $\kappa t=\pi$, pericenter). The diagram shows the radii of a flat rotation curve (V=const., $\Omega \sim 1/R$, and $\kappa = \sqrt{2}\,\Omega$). Note that the motion of the star can also be described as a rotation at velocity $\Omega$ along an ellipse with precession rate $\Omega-\kappa/2$.

This rapid variation of $\Omega \sim 1/R$ is the origin of the winding of the arms. If the waves rotate around the center with angular velocity $\Omega_p$ close to the precession rate $\Omega-\kappa/2$, then even a small force resulting from the self-gravity of matter will force the arms to rotate together at the same speed.

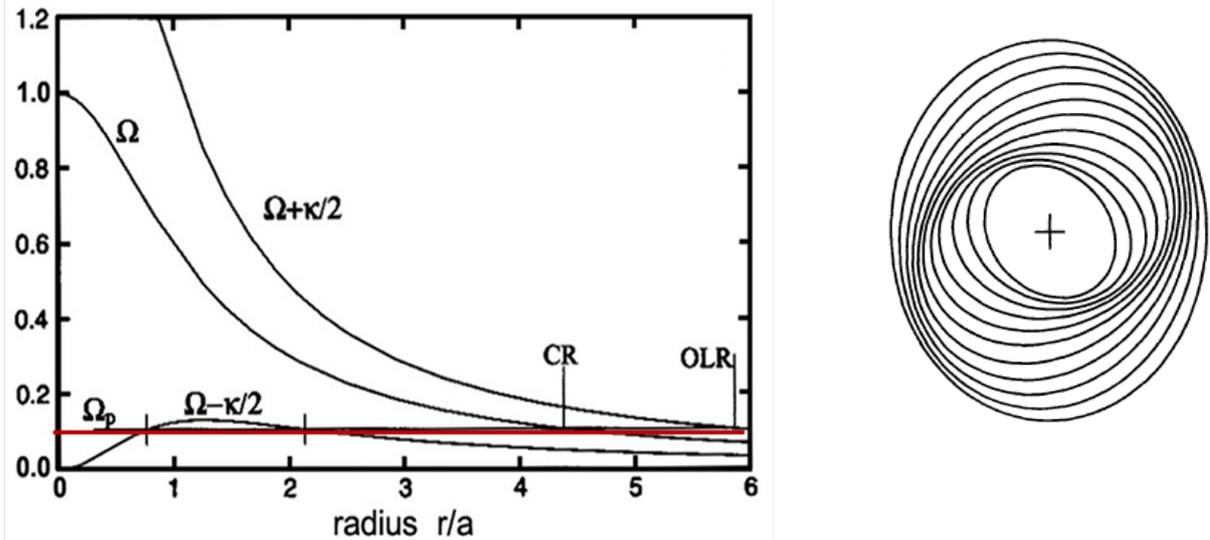

**Figure 12: Left**: Frequencies of the rotation curve of a typical galaxy, similar to the Milky Way. The precession rate of the elliptical orbits $\Omega-\kappa/2$ is almost constant, which suggests that waves rotating at this velocity $\Omega_p$ will not wind whenever self-gravity causes them to retain their coherence. The intersection of $\Omega_p$ with $\Omega$ is the corotation (CR), and $\Omega-\kappa/2$ is the outer Lindblad resonance (OLR). **Right:** Distribution of a few elliptical orbits precessing by $\Omega-\kappa/2$. Initially, each of them has a major axis that is rotated with respect to the previous one by a small angle that varies logarithmically with the radius. Superimposing these orbits creates a spiral, describing the concept of a density wave. Matter orbits along the ellipses and passes through two arms on each rotation. Reproduced from Kalnajs (1973).

### 3.2 The theory of density waves

We therefore need to show that self-gravity can force the precession rates to remain confined around a shared value. To do this, Lin & Shu (1964) had the idea of developing the equations of tightly wound waves in the WKB ("Wenzel, Brillouin, Kramers") approximation. The wavelength $\lambda$, i.e. the space between two arms, is assumed to be small relative to the radius R of the galaxy, so that $\lambda \ll R$. This has the advantage of allowing a local approximation (the effects of distant waves cancel out). Without this approximation, the equations would be impossible to solve analytically.

To find the wave dispersion relation, we assume a sinusoidal perturbation in the disc surface density $\Sigma = \Sigma_0 + \Sigma_1(r) \exp[-im(\theta-\theta_0) +i\omega t]$, where $(r, \theta)$ are the polar coordinates in the plane of the disc, m is the number of arms, and $\omega$ is the temporal frequency at which the arm is crossed ($\omega/m$ is the angular velocity). For small perturbations, we linearize the dynamical equations, namely the Poisson equation, which relates the density $\Sigma$ to the potential, and the collisionless Boltzmann equation, which governs the distribution function of stars in the potential. The stars of the galaxy may be assumed to be collisionless because the time required to exchange energy through 2-body collisions would be 10 million times the age of the Universe.

These equations incorporate the shape of the spiral waves, and in particular the "pitch" angle, namely the angle between the spiral and the circle of radius r. This pitch angle i satisfies $\tan(i) = 1/r \, dr/d\theta_0 = 1/(kr)$, where k is the wavenumber, $k = 2\pi/\lambda$.

If we assume that the spirals are very tightly wound (WKB approximation), the pitch angle is close to zero, and kr >> 1, or λ << r. There are two possible winding directions, with either "trailing" or "leading" waves, as shown in Figure 13.

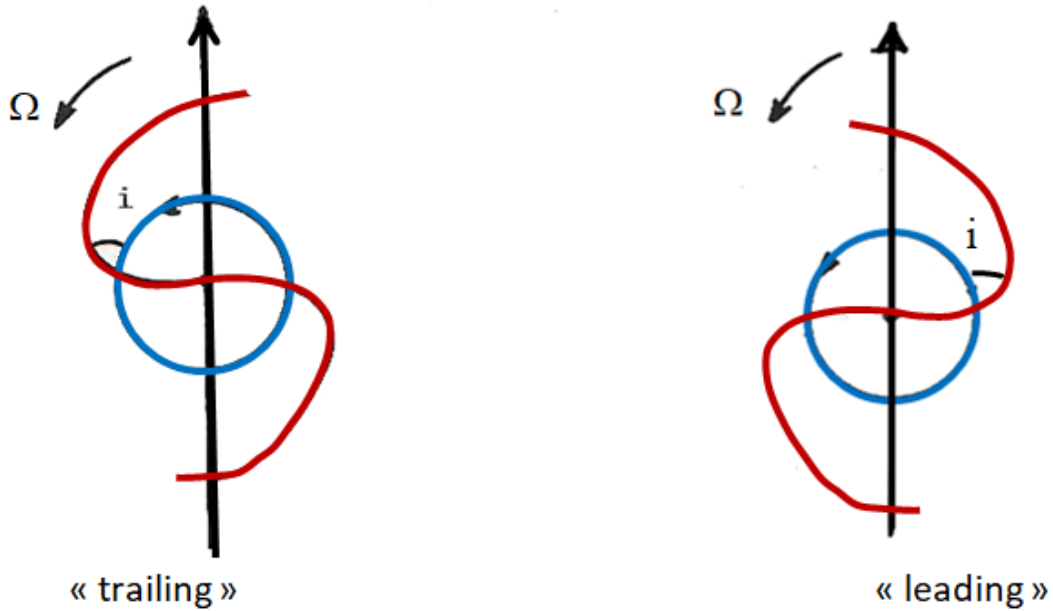

**Figure 13:** Winding direction of the spiral arms relative to the direction of rotation of the galaxy. On the left, matter enters through the concave side of the arm, and the spiral is "trailing." On the right, matter enters through the convex side, and the spiral is "leading."

The relation found by Lin & Shu (1964) shows the existence of two branches, short waves and long waves. The WKB approximation suggests that only the short branch is consistent with the hypotheses. Waves can only propagate everywhere in the disc, between the two Lindblad resonances (inner and outer), if the disc is not too "hot," i.e. the Toomre parameter Q is only equal to 1. Otherwise, a prohibited zone develops around the corotation that is increasingly wide at high values of Q, as shown in Figure 14. In the figure, the frequency $\nu^2$ is plotted along the vertical axis as a function of the wavelength λ or wavenumber k=2π/λ. The frequency of the wave is the frequency at which matter encounters a spiral arm and can be written $\nu=m(\Omega_p-\Omega)/\kappa$ (normalized by the epicyclic frequency κ).

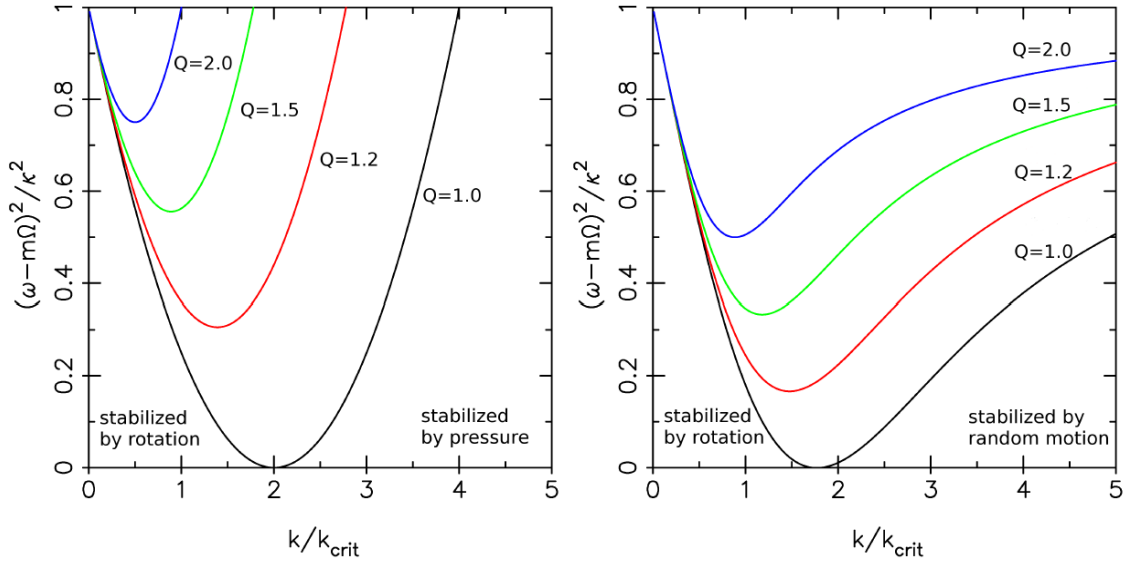

**Figure 14:** Wave dispersion relation, revisited by Dobbs & Baba (2014). The right-hand diagram corresponds to the stellar component, without collisions, and revisits the relation calculated by Lin & Shu (1964). The left-hand diagram is the interstellar medium, fluid with dissipation. The critical wavenumber is $k_{crit} = \kappa^2/(2\pi G\Sigma_0)$. The left part of the diagram $k \ll k_{crit}$, where $\lambda \gg \lambda_{crit}$, is the region of long waves; the right part, where $\lambda \ll \lambda_{crit}$, is the region of short waves. Normally, the WKB approximation imposes that only this branch should be considered. Nevertheless, the other branch gives a good idea of the solution.

Note that the vertical axis corresponds to the corotation at the bottom and the Lindblad resonances at the top, either inner (ILR) or outer (OLR). For $Q > 1$, there is a zone around the corotation that is impossible for waves to cross in this approximation (except by the tunnel effect). The dispersion relation can tell us about the geometric shape of the waves. The wavelength is proportional to Q (in the short branch) or ~1/Q for long waves. If the disc is heated by instabilities, the spiral winds in the long branch and unwinds in the short branch.

In reality, monochromatic waves are not physical, multiple waves can coexist together, and waves travel in wave packets with group velocity $v_g = d\omega/dk$. Furthermore, waves damp relatively quickly when the response of the gas is taken into account. Interstellar gas responds strongly to excitation due to its low velocity dispersion. Even though its mass fraction is relatively low compared to the mass of stars, its response is strongly non-linear and dissipative. There are shock waves at the entry to the spiral arms. The density contrast in the gas is much stronger, around 5-10, compared to just 1.2-1.5 for stars. The gas is compressed when it enters the arms, which triggers the formation of new stars. This makes the arms bluer and brighter, even though the density contrast of the stars is not very high. The shock waves create large velocity variations as they pass through the spiral arms – the characteristic "streaming" motion of density waves.

Ultimately, the question of the displacement of wave packets with sufficiently high velocity and the damping of waves as they cross the disc reduces the winding problem of physical spiral arms to another problem: the persistence of spiral density waves. What mechanism is at the origin of the formation of these waves, and, after being excited by an interaction or internal instability, how can the persistence of the spiral waves be explained?

Part of the solution is based on the reflection of waves, either at the center of the galaxy, or at the resonances, combined with an amplification effect. The waves either travel to the center of the galaxy, where they are reflected, or they are reflected before reaching it by a heat barrier if the Toomre parameter Q is too high in the center. The corotation plays a special role because the waves have negative energy travelling inward and positive energy travelling outward. At the corotation, a wave travelling outward can be partially transmitted and partially reflected inward. Transmitting a wave with positive energy outward amplifies the wave with negative energy that is reflected inward.

The swing phenomenon can cause a spectacular amplification that was studied and described by Toomre (1981). The amplification process occurs when a leading wave packet transforms into a trailing wave packet. It is caused by several mechanisms whose effects align and act concertedly: the differential rotation of the galaxy, self-gravity, and epicyclic motion, which operate like a swing that allows self-gravity to act for longer.

The tidal forces caused by interacting with a companion were suspected to be responsible for the excitation of spiral waves as early as the 1970s (Toomre & Toomre 1972). Tidal forces have the right symmetry to form two spiral arms, and most galaxies with coherent spirals do in fact have two arms. Indeed, the azimuthal dependency of the tidal forces is a function of $\cos(2\theta)$, which explains the perturbation with m=2. However, tidal forces mostly act at the edges of the galaxy, with little effect on the center. It can be shown that these forces vary like the square of the distance to the center for a distant companion. In the prototypical case of Messier 51 and its companion NGC5195, which is located at the end of its spiral arm (cf. Figure 10), the spiral structure is very pronounced immediately from the center of the galaxy where the tidal forces are acting. And indeed, in the very simple simulations by Toomre & Toomre (1972) without self-gravity for a restricted 3-body problem, the tidal spiral arms only form at the edges, and spirals are not created at the center. Self-gravity is required to reproduce the process of swing, and the characteristic amplification generated by self-gravity combines with the swing from the epicycles. The waves then propagate very quickly toward the center and are amplified so much that the tidal effects are much stronger inside the disc than at the edges, unlike in the test particle simulations.

Note that, if the orbit of the companion is inclined with respect to the plane of the first galaxy, the vertical tidal forces can distort the plane into a pancake shape. This is described as "warping." In this case, the azimuthal symmetry of these forces is m=1.

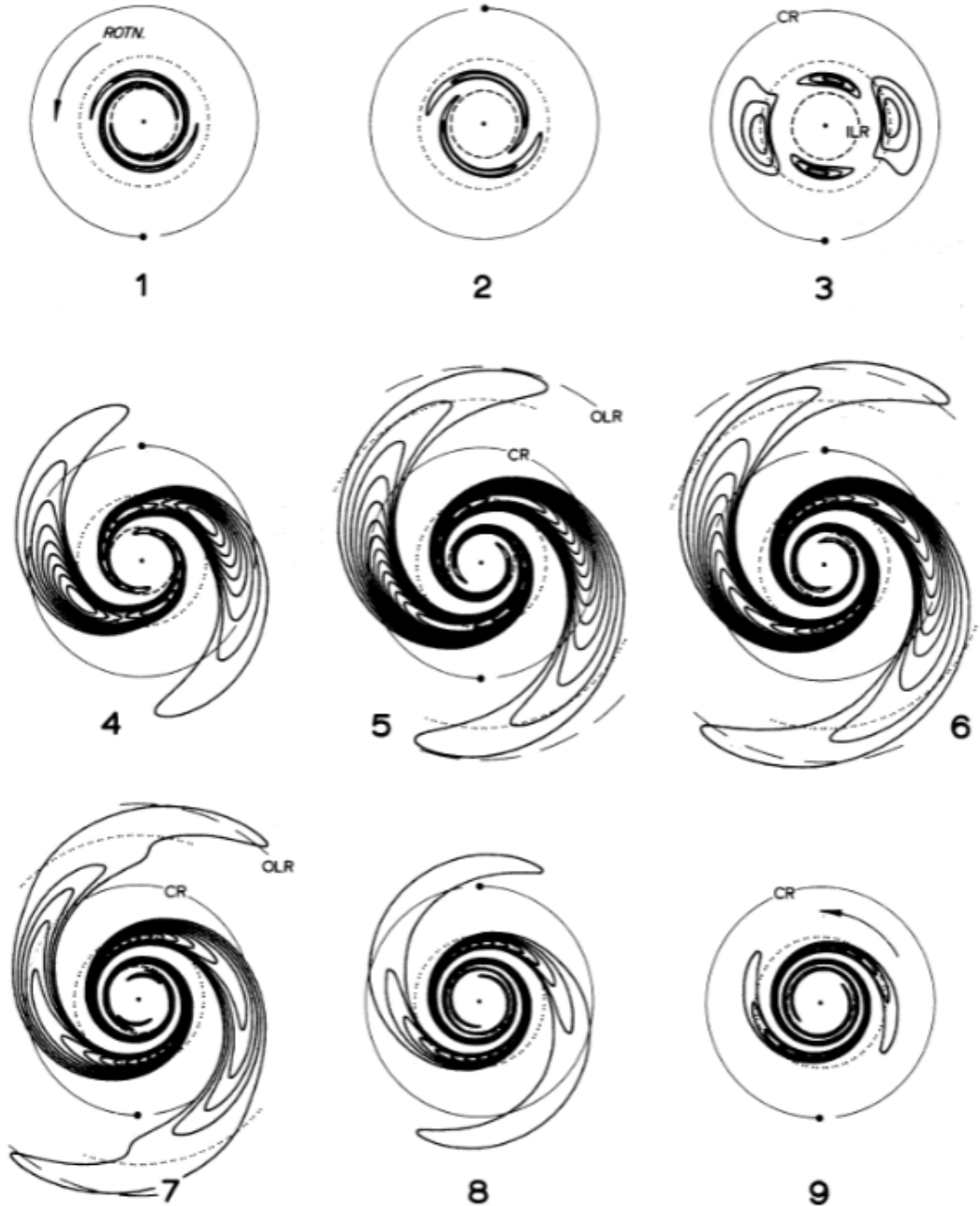

**Figure 15:** Illustration of the amplification of the swing. In this simulation, a leading wave packet develops during the first three epochs, then unwinds and transforms into a trailing wave packet. Its amplitude increases considerably under the combined effect of self-gravity, differential rotation, and epicyclic swing. The wave packet then propagates toward the center, where it is damped. Reproduced from Toomre (1981).

## 3.3 Role of gas and star formation

When passing through the spiral arms, which form a gravitational potential well, the gas accelerates and is subjected to shock waves. The compression favors the transformation of the diffuse atomic gas HI into the denser molecular gas $H_2$. Thus, dust and molecular gas follow a very narrow spiral inside the spiral arm (cf. Figure 16).

The ionized regions (which emit the recombination line Hα) indicate young and massive stars whose lifetime is less than the time required to pass through the spiral arm. However, the images have various irregularities, and the arms appear somewhat chaotic. Star formation does not always proceed in an orderly fashion, and there are frequent exceptions, with stars forming between the arms, branches, and extensions.

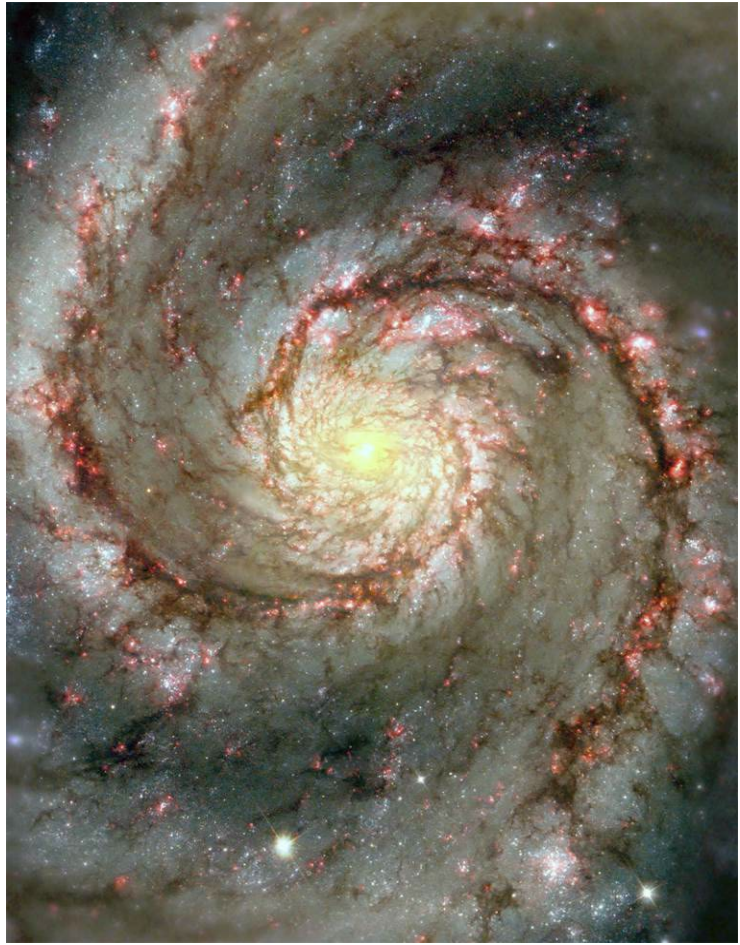

**Figure 16:** Multicolor image obtained by the Hubble Space Telescope. The regions of ionized gas that emit the Balmer Hα line of atomic hydrogen are visible in pink. Only very hot stars of types O and B are capable of ionizing the surrounding gas. These massive stars consume their gas fuel very quickly, with a typical lifetime of 10 million years. They tend to be created at the entry of the spiral arms and expire before the crossing is complete. As a result, the arms are very bright and pink. Credit NASA/ESA.

This is all caused by the non-linearity of the physics of gas, star formation, and feedback processes. Simulations have shown that the gas in the arms is subject to shearing (or Kevin-Helmholtz) instabilities that can cause extensions shaped like "spurs," as is often observed in the spiral arms of galaxies (cf. Figure 17).

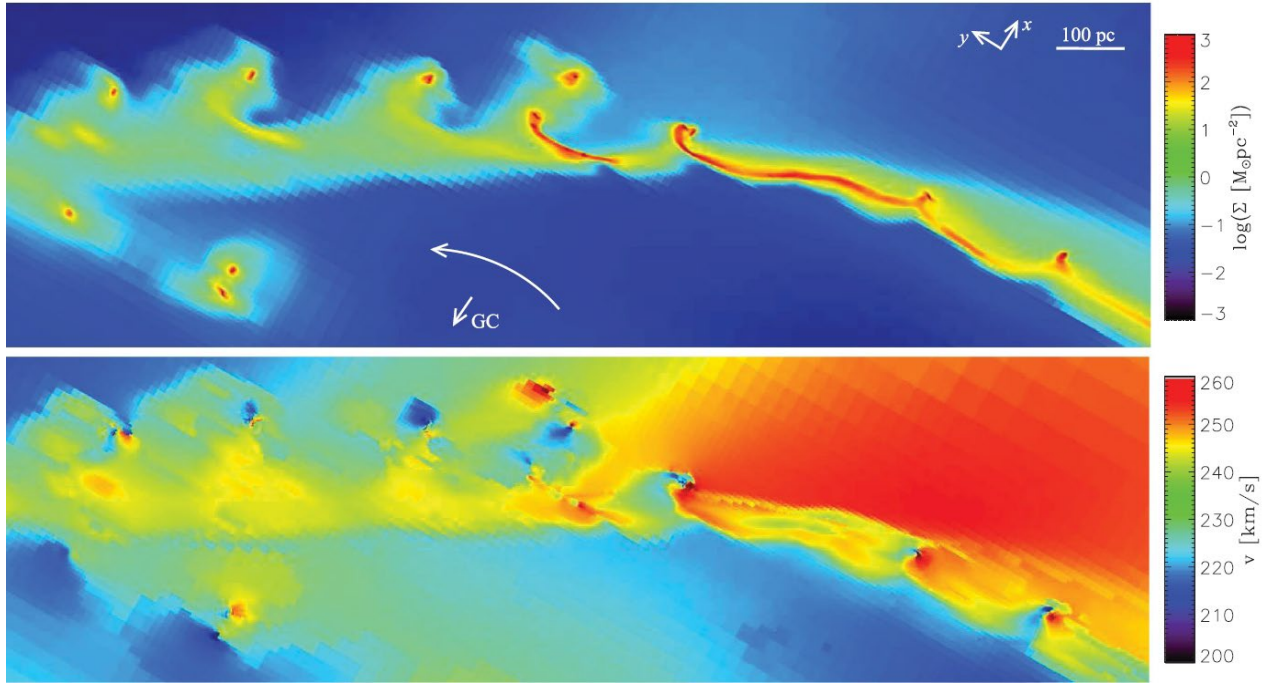

**Figure 17:** High-resolution simulation of the gas in a spiral galaxy, with the associated star formation and feedback. The top image shows the surface density of the gas. The galaxy center is marked by a downward white arrow. Instabilities can be seen to develop, forming extensions and spurs. The bottom image shows the velocity field for the same region of the spiral arm. The very high velocity gradient in the right section produces a shear; this is the source of the Kelvin-Helmholtz instabilities developing on the left. Reproduced from Renaud et al (2013).

## 4  Bars: drivers of evolution

Two thirds of spiral galaxies are barred, classified as SAB or SB. Of all disc galaxies, around one third are strongly barred (SB) and one-third are more moderately barred (SAB). Bars favor coherent spiral or "grand design" structures. Interactions between galaxies are also a determining factor, as described above. Flocculent spiral galaxies are therefore less common among barred galaxies, as shown in Figure 18.

The discs of galaxies are perpetually evolving; in extreme cases, they may be destroyed and reform. Whenever a galaxy evolves under its own gravity, it tends to concentrate its mass to approach the state of least energy. But rotation counteracts the concentration of mass. We must therefore determine the mechanisms to evacuate the angular momentum outward. The disc of a galaxy plays the role of an accretion disc. It is there to transfer angular momentum out of the galaxy.

Wherever there is interstellar gas, viscous torques are a possible candidate for this transfer. Dissipating the energy of this gas does indeed reduce its velocity dispersion, and the gas remains a "cold" component, but viscous torques are insufficient. The gas is very diffuse and highly fragmented into clouds. Dense molecular clouds do not behave like a fluid but like ballistic particles that are only involved in a collision once or twice per orbit around the galaxy center.

Gravity torques are therefore necessary to transfer the angular momentum. Asymmetries such as spiral arms or bars created tangential forces and torques. However, spiral arms are more transient, and their action is less impactful than bars. Bars are quasi-stationary density waves that can be viewed as a combination of leading and trailing wave packets. They are robust but still evolve secularly.

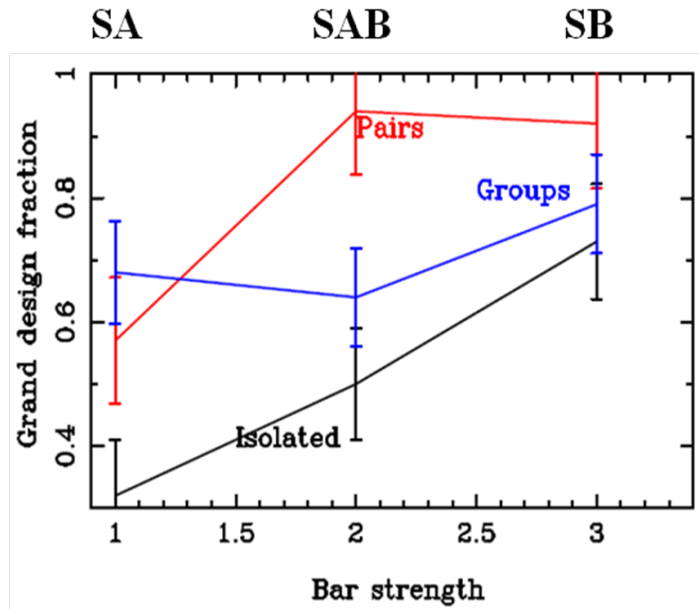

**Figure 18:** The fraction of coherent spiral or "grand design" structures as a function of the bar strength for three types of environment: isolated galaxies, groups of galaxies, and pairs of galaxies. Barred galaxies are much more likely to have a spiral structure with density waves than a flocculent structure. Reproduced from Elmegreen & Elmegreen (1983).

## 4.1 Formation of bars

The goal of the first self-gravity simulations of galaxy discs in the 1970s and 1980s was to test the theory of density waves. But, although the simulated discs all started at equilibrium, satisfying the Toomre stability criterion, Q>1, they were all unstable and ultimately always formed bars (Hohl 1971, 1976). The instability of bars had not been foreseen by the theory of density waves, since the latter had been working with the hypothesis of tightly wound waves, whereas bars are on the contrary very open waves.
Several solutions have been considered to explain the stability of discs, since at least one third of spiral galaxies are not barred. The idea is to make the disc more turbulent and unsuitable for the development of waves, with a Toomre parameter that is Q>2 or higher near the center and decreases away from it. Recall that the criterion Q=1 stabilizes against disc fragmentation by means of velocity dispersion at small scales and differential rotation shearing forces at large scales. The first effect of increasing the value of Q is to increase the velocity dispersion. But another way to achieve stabilization is to make the disc less self-gravitating by placing it within a spherical potential from the dark matter halo. This idea was proposed by Ostriker & Peebles (1973). At the time, the rotation curves of spiral galaxies were beginning to show that invisible matter necessarily exists in the outer regions.

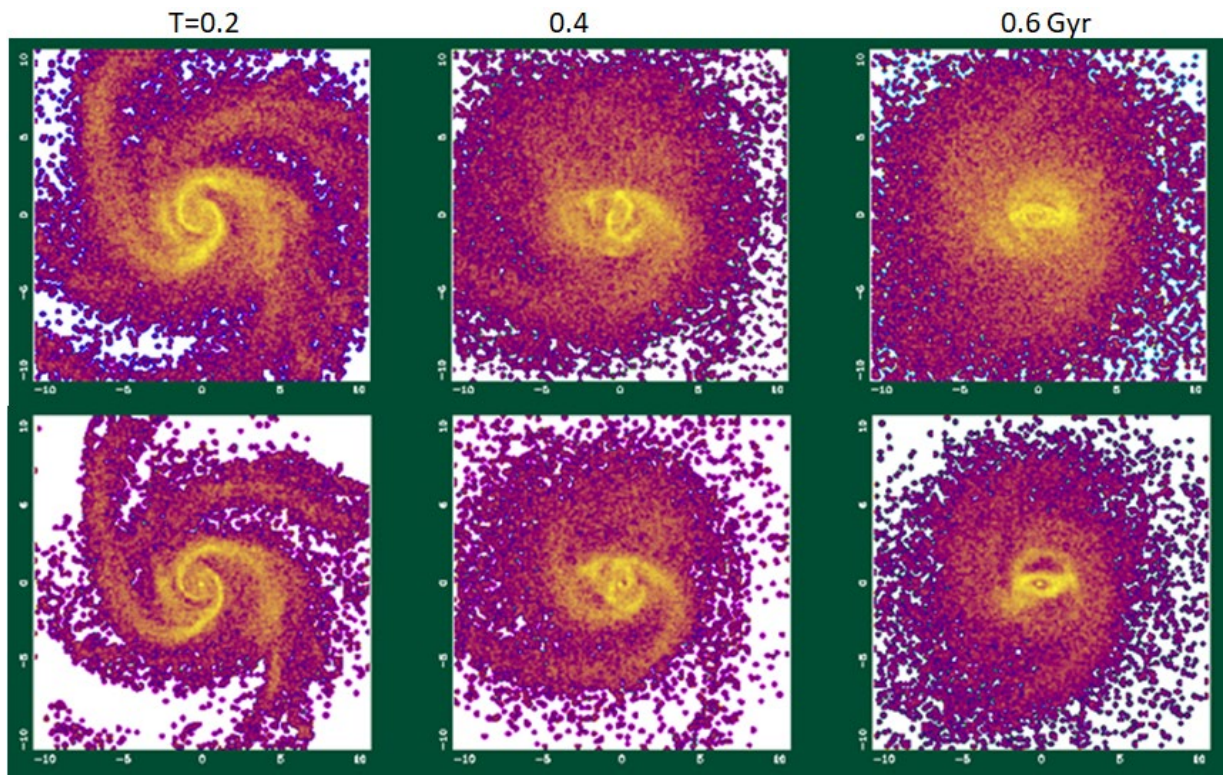

**Figure 19a:** Simulations of a self-gravitating disc, containing stars (top panel) and interstellar gas (bottom panel). Both components are subject to the same self-gravitating potential well, but they behave very differently, since the gas is dissipative and is constantly cooled by radiation, whereas there is nothing to counteract the heating of the stars. Each epoch lasts for 200 million years. At first, a transient spiral forms to evacuate the angular momentum.

Ostriker & Peebles demonstrated stabilization for dark matter halos by performing self-gravitating N-body simulations with N=300! They chose a spherical halo that was simulated non-coherently. An analytic potential was applied to the disc without feedback. The impact of this stability demonstration on the astronomy community was significant; it proved the utility of dark matter halos and gave evidence of their existence at a time where a consensus had not yet been reached. Later, once coherent simulations of dark matter could be performed with particles, like for stars and gas, it was observed that the dark matter interacts with the disc by exchanging angular momentum (cf. Athanassoula 2002). By accepting angular momentum from the disc, it enables a transfer and allows the bars to develop. Instead of suppressing bars, dark matter halos facilitate them!

Since then, various evolutions have improved the realism of simulations. First, interstellar gas was incorporated, a dissipative component that behaves differently than stars, which are a collisionless medium (cf. Figure 19). Later, incorporating star formation from gas and its feedback allowed various stellar populations in the disc to be accounted for, as well as stochastic effects in the morphology of the spiral arms. Resonance effects allowed the formation of rings to be explained (Buta & Combes 1996, cf. Chapter 1), and radial migration (Sellwood 2014) shed light on metallicity gradients.

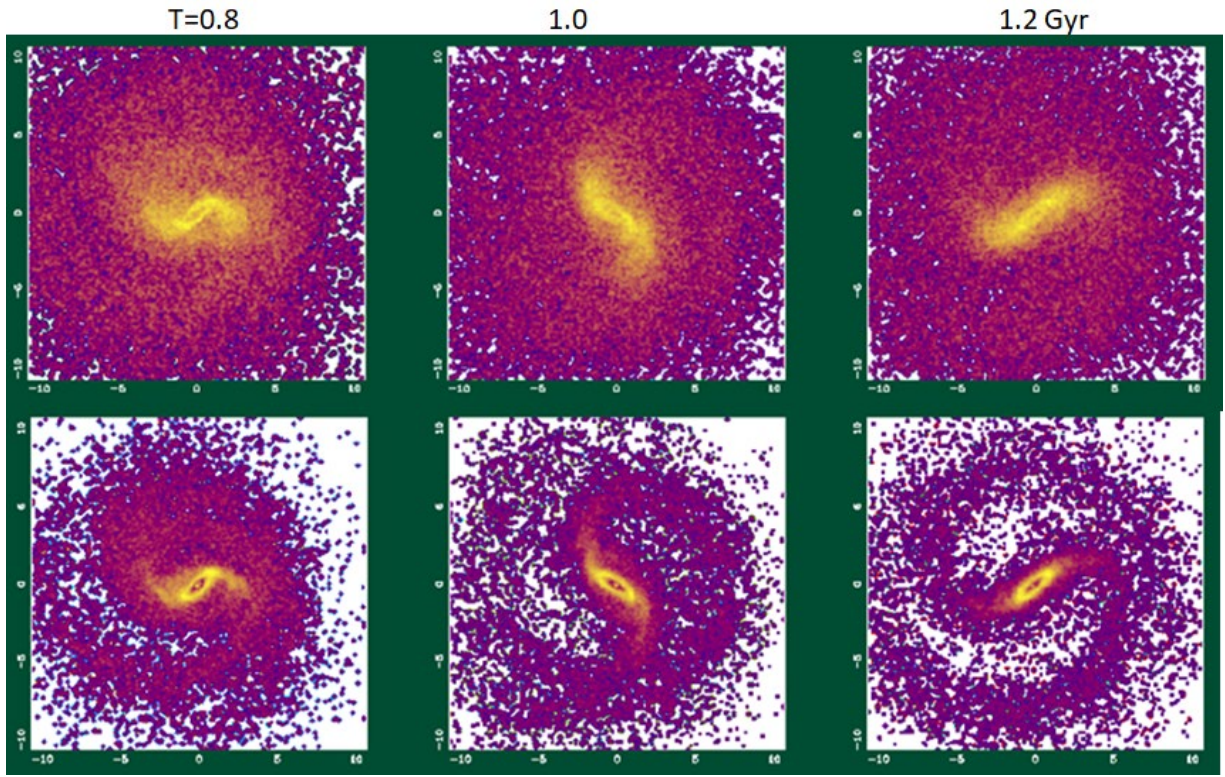

**Figure 19b:** Continuation of the simulation of the formation of a bar. A barred wave develops in the stars, while continuing to generate spiral waves in the gas. Note the resonance rings in the gas at the inner Lindblad resonance. The ring has an elliptical shape that is elongated parallel to the bar.

## 4.2 Orbits in a barred galaxy

To better understand the dynamical phenomena associated with bars, studying the orbits of stars in a barred potential, in particular in a frame of reference that rotates with the bars, is essential. These orbits were calculated and precisely described by Contopoulos & Papayannopoulos (1980). In the frame rotating with the bar at angular velocity $\Omega_b$, the inertial term $½\, \Omega_b^2\, r^2$ is added to the bisymmetric gravitational potential $\Phi$ (m=2) of the bar (since the inertial force that derives from this potential is $\Omega_b^2\, r$). Hence, the equivalent potential $\Phi$ eq is given by $\Phi$ eq = $\Phi$ (r, θ, z) - $\Omega_b^2\, r^2/2$.

The energy in this frame, or the Jacobian EJ, is an integral of motion, since the potential is independent of time:

$$EJ = v^2/2 + \Phi (r, \theta, z) - \Omega_b^2\, r^2/2.$$

However, the kinetic momentum Lz perpendicular to the plane is not conserved, since the potential is no longer axisymmetric; the bar introduces tangential forces and gravity torques. Considering the shape of the equivalent potential in the rotating frame is instructive because it determines the shape of the orbits. Figure 20 shows the equipotentials. The bar is horizonal, and the Lagrange points are the stationary points: L4 and L5 are maxima, L1 and L2 are saddle points, all around the corotation with the bar.

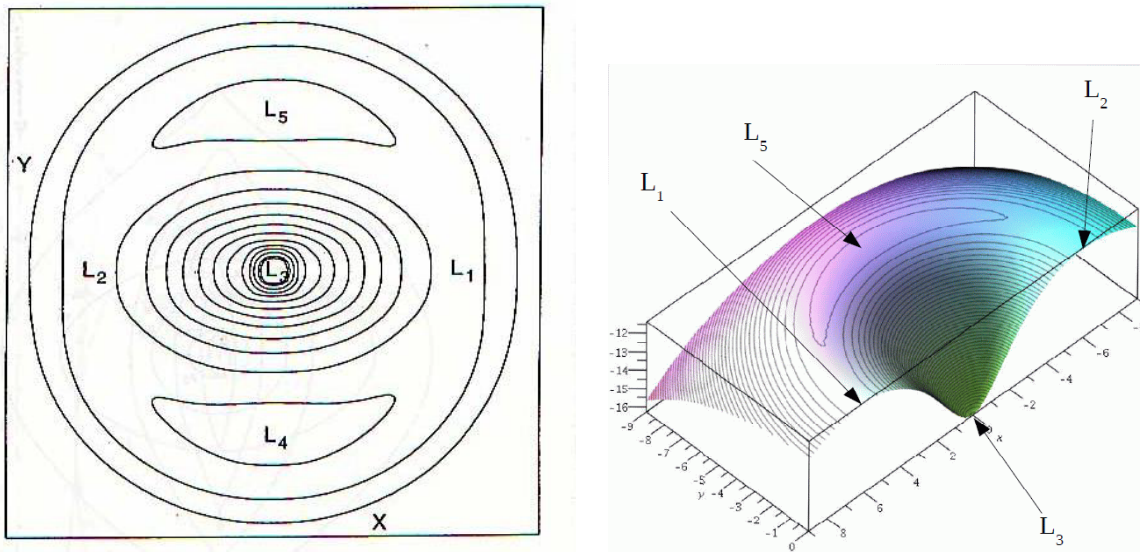

**Figure 20:** Left: shape of the equipotential lines in the frame that rotates with the bar, for a bar along the Ox axis. There are 5 Lagrange points: L3 at the center of the potential, L1, L2, L4, and L5 at the highest points, located at the corotation of the bar along the axes Ox and Oy. Right: perspective and cross-section to better illustrate the potential in three dimensions. Credit D. Pfenniger.

The shape of the equipotentials already suggests that the principal orbits will align with the bar inside the corotation; outside, they will be perpendicular to it. There will be stable orbits that only revolve around L4 and L5. The periodic orbits form a skeleton that defines the orbits of every star in the galaxy. They attract and trap the other orbits, the majority of which are not periodic. There are also chaotic orbits that are not trapped around periodic orbits.

More precisely, the orbits are divided into the following families:

(1) Very close to the center, and up to the corotation if there is no inner Lindblad resonance (ILR), the orbits are parallel to the bar and support it. This family is called x1.

(2) Between the two inner Lindblad resonances, if they exist, there are x2 orbits that are perpendicular to the bar, direct, and stable. The x2 family disappears if the bar strength is too high (which also removes the ILRs).

(3) If there are 2 ILRs, the x1 orbits parallel to the bar resume between the second ILR and the corotation.

(4) At the corotation, there are stable orbits around the Lagrange points L4 and L5 that do not rotate, in the reference frame of the bar.

(5) After the corotation, the orientation of the orbits changes again, and they become perpendicular to the bar and do not support it. The bar must therefore end at its corotation (cf. Figure 21).

There is a simple way to summarize the orientation of the orbits: they are either parallel or perpendicular to the bar, and the orientation of the orbits changes (by 90°) at each resonance.

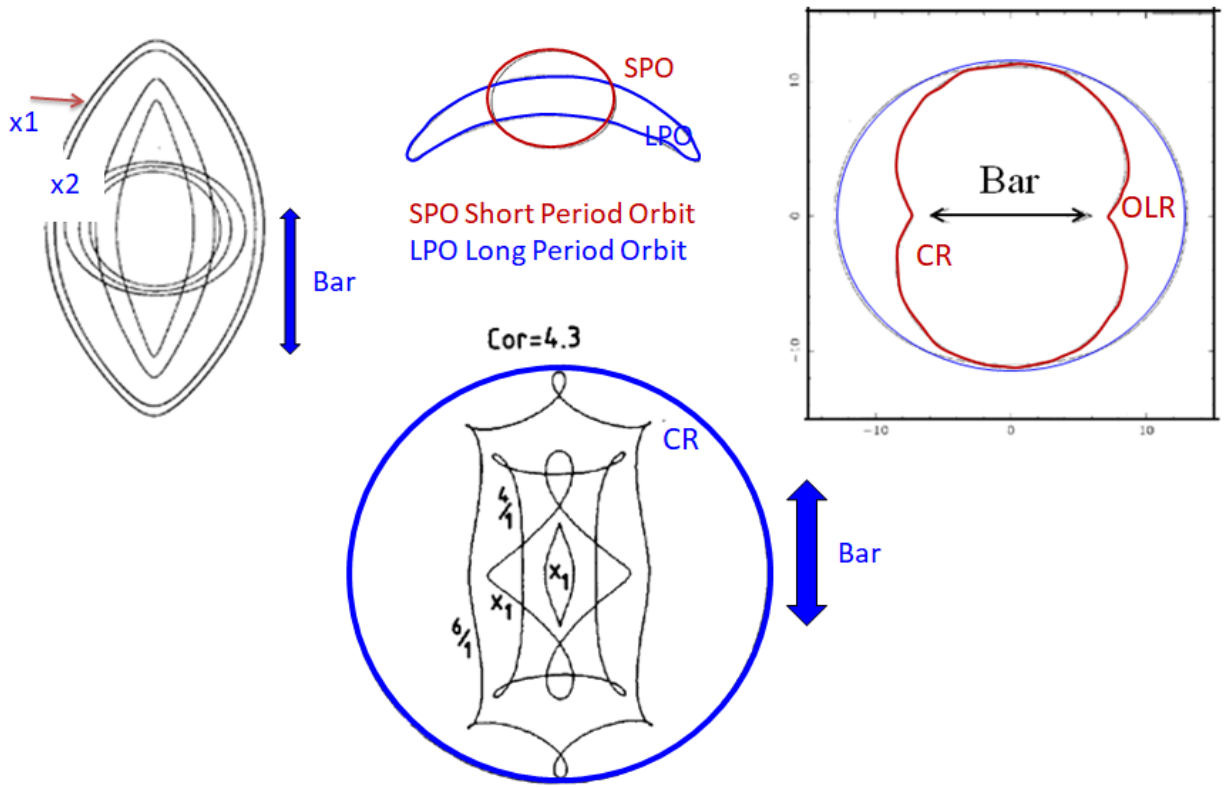

**Figure 21:** Periodic orbits characteristic of a barred galaxy in the rotating frame of reference. Left: the x1 and x2 families at the center of the bar. Center: the stable orbits around the L4 and L5 Lagrange points. Right: periodic orbits perpendicular to the bar, outside of the corotation, up to the outer Lindblad resonance (OLR). Bottom: higher resonances, very close to the corotation. Reproduced from Contopoulos & Papayannopoulos (1980).

By knowing the families of orbits, we can understand the evolution of the bars better. When bars first form, simulations show that they have relatively high angular momentum, and no inner Lindblad resonance. The orbits in the x1 family support the bar, which grows as it traps more and more stars with a low precession rate, slowing down. One ILR can potentially develop, followed by a second, and orbits in the x2 family can weaken the bar, possibly even leading to its destruction. This process can be viewed as a form of self-regulation of the bar strength.
Furthermore, the orbits no longer support the bar past the corotation (CR), becoming perpendicular to it. This explains why bars generally end just inside their corotation, sometimes with a 4:1 resonance ring not far from the CR. This is an excellent diagnostic for determining the angular momentum of the bar $\Omega_b$.

## 4.3 Response of gas to a barred potential

In the disc of a barred galaxy, the gas tends to follow the same periodic orbits as the stars; however, stars are collisionless, whereas interstellar gas is not. The gas clouds collide, which prevents the orbits or streamlines from crossing in practice. Instead of rotating abruptly by 90° at each resonance, the gas streamlines turn gradually to finally reach 90° at each resonance. The gas response therefore has a spiral structure, whereas this spiral disappears with stars (cf. Figure 22).

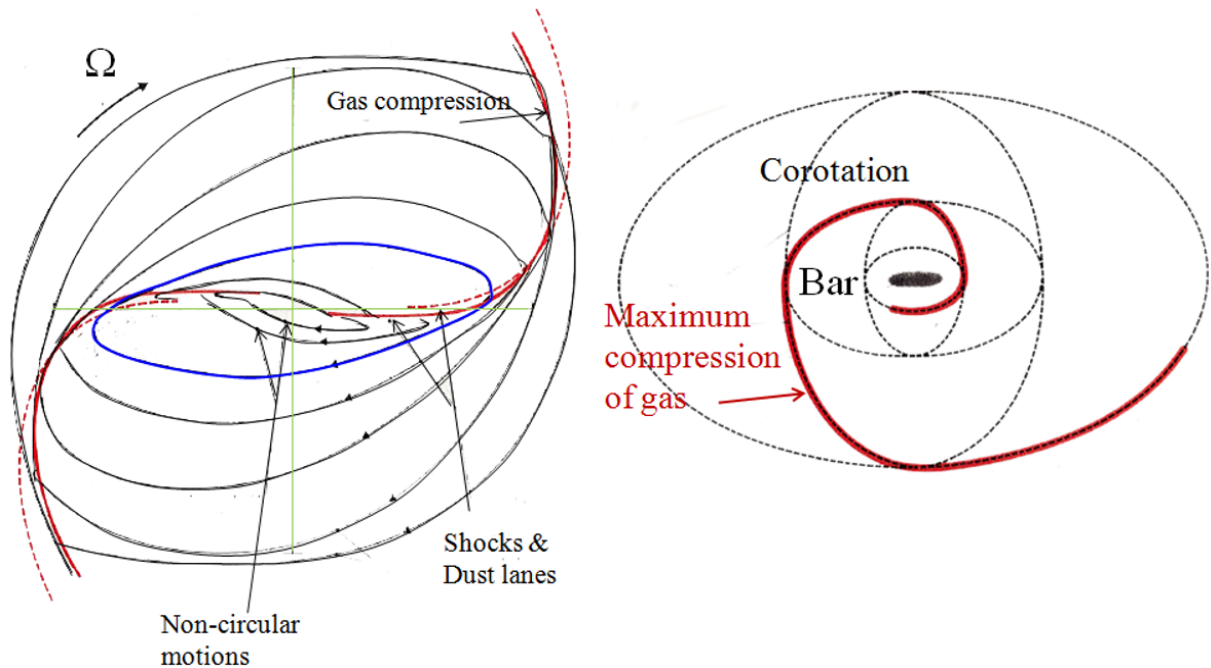

**Figure 22:** Due to collisions between gas clouds, the streamlines of the interstellar gas gradually rotate by 90° at each resonance, rather than abruptly like for stars. The orbits resemble Kalnajs kinematic ellipses (cf. Figure 12), and they produce spiral arms as they tighten. If there are three resonances in the disc, the spiral winds by 270° (cf. right diagram).

Once the gas has distributed itself into a spiral structure, it experiences tangential forces from the bar, and hence torques, which causes a transfer of angular momentum. This does not happen if the gas is distributed axisymmetrically (e.g. in a ring) or symmetrically with respect to the bar (e.g. aligned with the bar). In this case, the tangential forces and associated torques cancel out on average over each orbit. Figure 23 gives a geometric representation showing the direction in which the torques act. The bar defines four quadrants in the space of the disc: two opposite quadrants (NE and SW) where the torque is positive, and two other quadrants (NW and SE) where the torque is negative relative to the direction of rotation of the galaxy. The torques change sign at each resonance. Outside of the corotation, the gas receives angular momentum and is driven outward to the outer Lindblad resonance (OLR), where it forms a somewhat elongated ring (perpendicular to the bar just before the OLR, in the shape of a figure of eight; or elongated parallel to the bar just outside of the OLR, cf. Figure 21). Arranged as a symmetric ring relative to the bar, it no longer receives torque and is stable. Inside of the corotation, the gas loses angular momentum and spirals toward the center. If there is an ILR, the gas stops there and forms a nuclear ring. Inside the ILR and up to the center, the spiral continues in the "trailing" direction in opposite quadrants, and the torque is positive, carrying the gas toward the ILR, where it accumulates and forms stars. If there are two ILRs, and especially if the gas falls within the sphere of influence of the central black hole, transient leading spirals can form rapidly in regions where the precession rate $\Omega - \kappa/2$ increases as the radius decreases. In these cases, the sign of the torque is once again inverted, and the gas can supply the nucleus (cf. Buta & Combes, 1996).

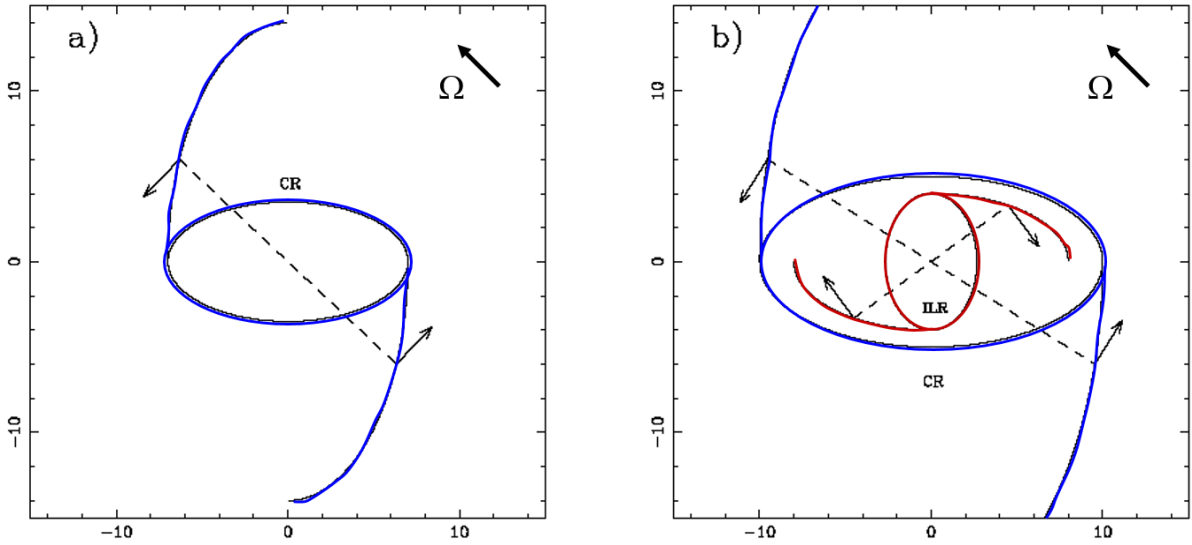

**Figure 23:** Diagram showing the direction of the gravity torques exerted by the bar on the gas distributed in a spiral structure: a) outside of the corotation, the torque is positive, and the gas is driven out toward the outer Lindblad resonance (OLR); b) inside of the corotation, the torque is negative, and the gas will spiral toward the center.

After the gas loses its angular momentum, it accumulates toward the center and forms stars. The mass concentration gradually increases, which also increases the precession frequency $\Omega - \kappa/2$ and generates two ILRs. The bar is weakened by the perpendicular orbits, which causes a secondary or nuclear bar to decouple within the last ILR. This bar, affected by the high precession rate at the center, rotates with a higher angular velocity than the primary bar. In numerical simulations, it is often possible to observe a nuclear bar that shares resonances with the primary bar in order to minimize chaotic orbits. For example, the ILR of the primary bar is the corotation of the secondary bar (Friedli & Martinet 1993, Garcia-Burillo et al 1998). The nuclear bars extend the action and gravity torques of the primary bar to the center of the galaxy. They supply the active galaxy nuclei (e.g. Shlosman et al 1989). There may even be a whole cascade of nested waves, as has been found in simulations (Hopkins & Quataert 2010).

**4.4  Vertical resonances and peanuts**

Viewing simulations of barred galaxies edge-on reveals a peanut structure, similar to a bulge but caused by the stars in the disc (Combes & Sanders 1981, Combes et al. 1990, Raha et al. 1991). These structures had been identified in the observations of the earliest catalogs but had previously not been interpreted. The peanut galaxy NGC 128, shown in Figure 24, is a prototype for them.

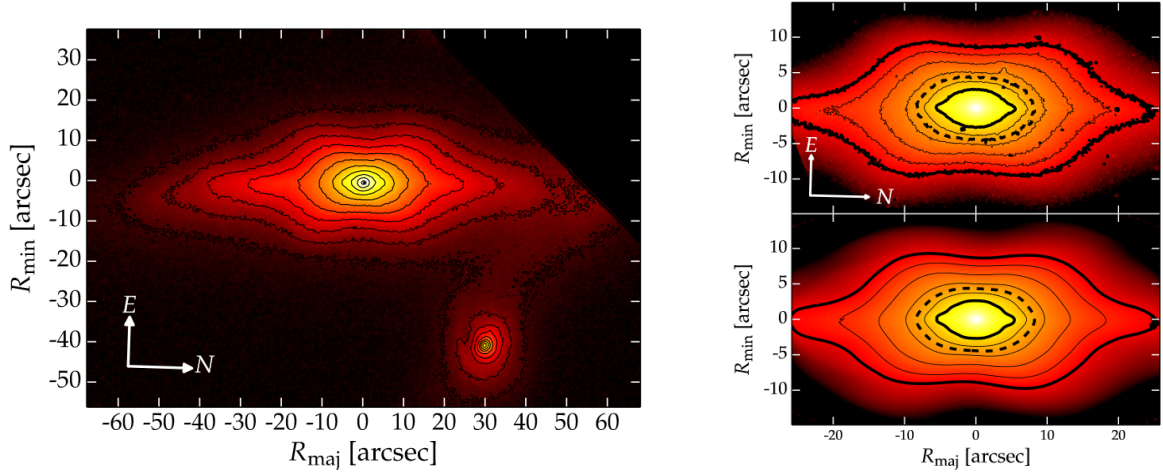

**Figure 24:** Example of a peanut structure: the galaxy NGC 128, viewed edge-on. On the left, red band image from the SDSS. On the right, near infrared image from the Hubble Space Telescope (top) and model with ISOFIT/CMODEL (bottom). Reproduced from Ciambur & Graham (2016). NGC 128 is in the process of interacting with its companion NGC 127 to the north-west (here, bottom-right).

Simulations have shown that this structure is caused by the vertical elevation of the stars of the disc, which resonate with the bar at the vertical Lindblad resonance radius when $\Omega_b = \Omega - \nu z/2$, where $\nu z$ is the vertical oscillation frequency (Combes et al. 1990). This resonance is not far from the inner Lindblad resonance (ILR) in the plane, since the frequency $\nu z$ is close to the epicyclic frequency $\kappa$. The peanut shape is one way to recognize the existence of a bar in galaxies viewed edge-on (Bureau & Freeman 1999). The peanut thickens the plane, which causes a three-dimensional dilution of the bar and weakens it. Nevertheless, the bar can still evolve by exchanging angular momentum with the dark matter halo. The bar strength may still increase, and even its angular velocity $\Omega_b$ may still decrease. Resonance then occurs at a larger radius, and a second peanut with a larger radius develops (Martinez-Valpuesta et al 2006).

The peanut shape is especially apparent when the bar is seen along its major axis in a galaxy viewed edge-on. In other projections, it mostly appears as a box shape. In any case, it is a bulge, called a "pseudo-bulge" as opposed to a classical bulge. The key differences between these two types of bulge are: their rotation – pseudo-bulges originate from the disc and are still strongly rotating, unlike classical bulges, which originate from minor mergers; their light profile – pseudo-bulges have a Sersic index close to n=1-2 rather than >4 like classical bulges; and their flattening – pseudo-bulges are flatter, whereas classical bulges are more spheroidal (Fisher & Drory 2010).

The bulge of our own galaxy, the Milky Way, is a pseudo-bulge caused by the vertical resonance of its bar in the center (e.g. Di Matteo et al 2015).

The orbits supporting the peanut shape of a pseudo-bulge have also been identified. They are either shaped like a "pretzel" (Portail et al. 2015) or a "banana" (Combes et al. 1990, and Figure 25).

- 32 -

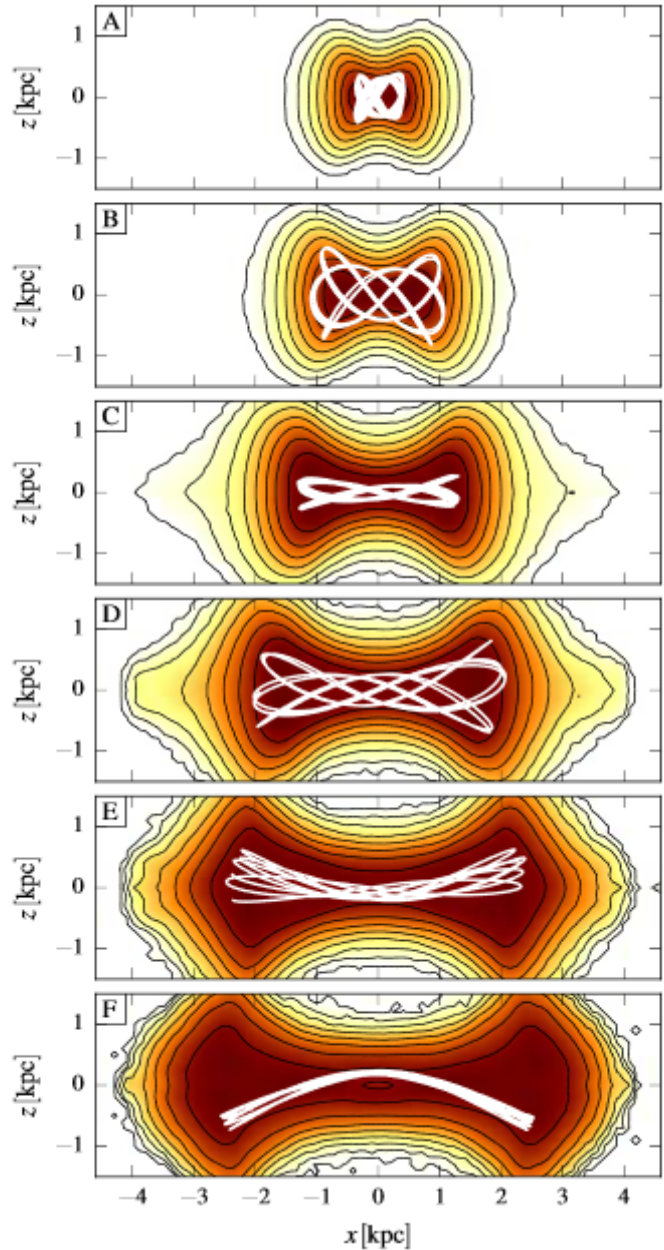

**Figure 25:** Families of orbits in the projection where the galaxy is viewed edge-on and the bar is viewed along its major axis. The orbit categories A to F make up the morphology of the peanut bulge. In particular, the category C contains "pretzels" and the category F contains "bananas." Reproduced from Portail et al (2015).

## 4.5 Dark matter and bars

In Section 4.1, we saw that dark matter halos facilitate the formation of bars by exchanging angular momentum. The presence of a bar cannot be stabilized by the dark matter halo, as predicted by Ostriker & Peebles (1973). Nevertheless, the bar can constrain the quantity of dark matter at the center of the galaxy, within the radius of the bar, by dynamical friction effects.

The standard cosmological model of dark matter (CDM) predicts a central concentration of dark matter, corresponding to a central peak in the radial density distribution, described as a cusp. This is namely the NFW profile (Navarro, Frenk & White 1997), where the density $\rho(r) \propto 1/r$, meaning that it satisfies a power law with slope 1. This cusp is not observed in rotation curves. Instead, the observed density is compatible with a core, i.e. a flat density function $\rho(r)$. The existence of a bar could further constrain the quantity of dark matter in the center. Indeed, as the bar rotates, it is slowed by dynamical friction on the dark matter halo, losing all of its speed if the halo density is too high. The corotation should therefore be pushed out of the disc, but this is not observed in practice. The relevant calculations were performed by Debattista & Sellwood (2000). They show that, to be compatible with the observations of rapid bars, the rotation curve must be dominated by the visible mass within the radius of these bars. The definition of a rapid bar can be specified in terms of the ratio of the corotation radius and the bar radius. The observed value of this ratio is RCR/Rbar ~1.2. The resulting constraint on dark matter is already significant, but it can be further reinforced by considering bars in galaxies with low mass or dwarf galaxies, such as the Large Magellanic Cloud. However, it remains difficult to find the pattern speed of the bar in these galaxies, which are irregular, very rich in gas, and do not have any resonance rings.

Another possibility might be to completely change the model by adopting modified gravity without a dark matter halo, for example the MOND model proposed by Milgrom (1983). A study of the formation and evolution of bars was performed by Tiret & Combes (2007, 2008). Without a halo of particles, dynamical friction is essentially inexistent for the bar, which does not slow down, as shown in Figure 26.

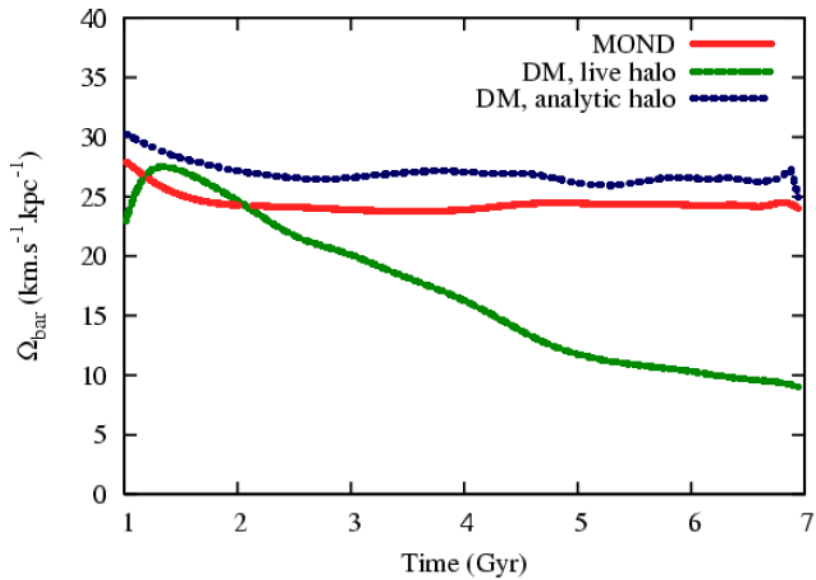

**Figure 26:** Evolution of the pattern speed of the bar over 7 billion years for 3 simulated models. In blue, a dark matter model (CDM) with a rigid dark matter halo that only imposes an analytic potential, without dynamical friction. In green, a realistic, live CDM model with particles, which slows the bar down by dynamical friction. In red, the MOND modified gravity model, without a halo of particles, with dynamical friction only from stars in the disc. Reproduced from Tiret & Combes (2007).

Note that the observed frequency of bars is also a very critical test for any dynamical model of galaxy formation, as well as any dark matter model. In a modified gravity model without a dark matter halo, the disc is completely self-gravitating and more unstable given the same observed velocity dispersion. We might therefore expect to find more unstable discs, either fragmented or forming more bars. But there are multiple effects combined: in the MOND model, the bars arise earlier and weaken faster by forming peanut bulges. Once they weaken, they can no longer be amplified, unlike the bars of the CDM model, where the halo can retake some angular momentum from the stars of the disc and amplify a second bar. In the modified gravity model, angular momentum is exchanged with stars in the disc, and the outer regions are stretched out. After the first bar weakens, these disc particles are no longer available to amplify a second bar. If we summarize and count the bars of both models, we find that they are slightly more numerous for MOND at a given strength, which seems to fit the observations better.

Even in the standard model, the frequency of bars and strong bars is problematic (cf. Block et al 2002, Eskridge et al 2002). For a sample of around 200 galaxies observed in the near infrared, the histogram of galaxies as a function of their bar strength is deficient for unbarred galaxies relative to simulated predictions. In simulations that account for gas, the bars lead the gas toward the center via gravity torque. This accumulation of matter in the center weakens the bar, especially since it accepted angular momentum from the gas by exerting torque upon it. This causes the bar to weaken even further, since it is a wave with negative momentum. To resolve this discrepancy between models and observations, we must consider that spiral galaxies are not isolated throughout their lifetimes but are enriched by accretion of matter originating from cosmic filaments. In particular, galaxies accrete gas that refills the disc and cools it down due to its low velocity dispersion. The disc then becomes unstable once again to form a second bar or reinforce the first bar (Bournaud & Combes, 2002, Bournaud et al 2005). In this way, by considering several barred episodes in spiral galaxies together, we can reproduce the near infrared observations of the frequency of bars in such galaxies more closely. For a good level of agreement, galactic discs would need to double their mass in 7-10 Gyr, i.e. less than the Hubble time. In the modified gravity model, the accretion of gas must also be taken into consideration, and the bar frequency is also compatible with observations (Tiret & Combes 2008).

## 5   Environment of spiral galaxies

### 5.1   Morphological segregation

As a population, spiral galaxies are found in different environments than the other categories of early-type galaxies, namely lenticular and elliptical galaxies. Dressler observed as early as 1980 that the fraction of the various populations varies highly significantly with the surface density of galaxies projected onto the sky (cf. Figure 27). A recent version of this diagram was compiled by Cappellari (2016) with more galaxies. Here, the main effect of the environment is not just to change the morphology of galaxies, but also to stop their star formation, transforming galaxies that are still active into inactive red galaxies. At the beginning of this chapter, we cited various mechanisms that can cause this phenomenon: stripping of gas by ram pressure, tidal interactions and harassment, and most importantly suffocation, by suppressing the accretion of cold gas from cosmic filaments. However, although it is conceivable to imagine that gas stripping could transform a spiral galaxy into a lenticular galaxy, the increased frequency of ellipticals due to mergers within galaxy clusters cannot be explained in this way, since the relative velocities within the clusters are far too high, with hyperbolic trajectories; galaxy pairs rarely form in rich environments. Within groups, merging pairs can still form, since the relative velocity is sufficiently low. It therefore seems that galaxy mergers resulting in elliptical galaxies take place at a lower level; they are pre-evolved outside of their clusters, so to speak, within a group or filament. These groups then merge within richer environments, namely clusters, supplying the latter with elliptical galaxies and explaining why such galaxies accumulate in denser regions.

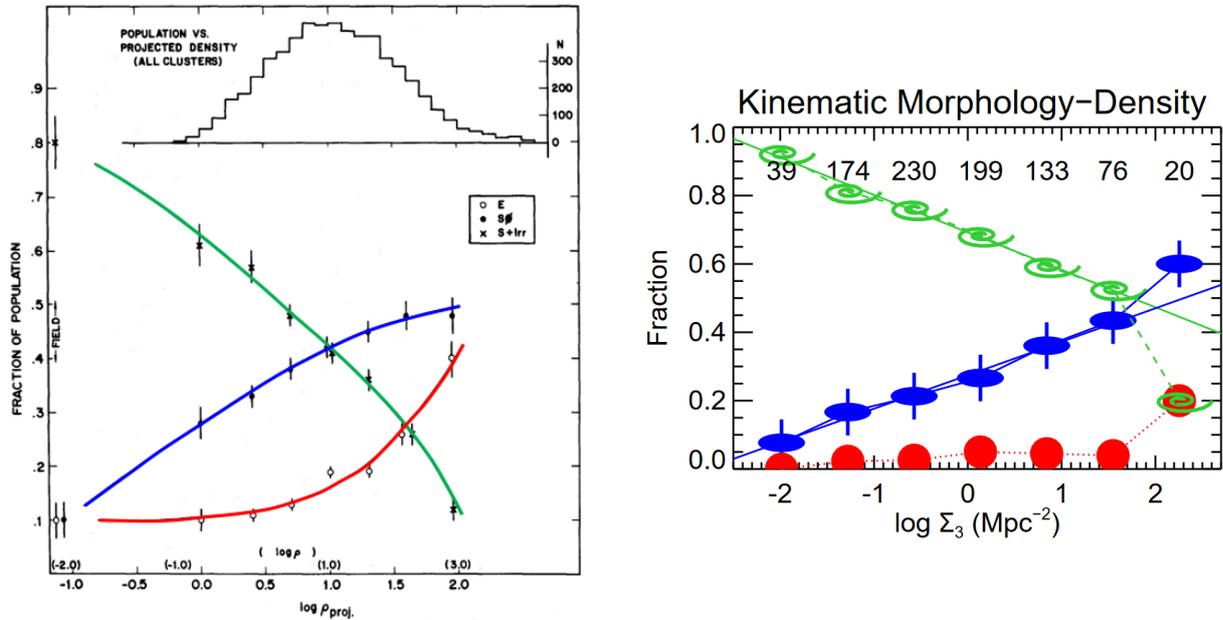

**Figure 27:** Fraction of galaxies for three different populations: spiral galaxies (green), lenticular galaxies (blue), and elliptical galaxies (red) as a function of the local environment, i.e. the surface density in galaxies $\Sigma$ or $\rho$-proj. On the left, the original diagram by Dressler (1980), with the number of affected galaxies shown above. On the right, a more recent version of this diagram, by Cappellari (2016).

## 5.2 The problem of bulgeless galaxies

Rich galaxy environments are fairly rare, and, locally, most galaxies are spiral galaxies (around two thirds). But the vast majority of them only have a very weak bulge, or no bulge at all. This observation is problematic for the standard cosmological model, which is hierarchical.

Locally, around two thirds of bright spiral galaxies have very little bulge or no bulge at all (Kormendy & Fisher 2008, Weinzirl et al. 2009, Kormendy et al. 2010). The remaining third consists of galaxies with both a classical bulge and a pseudo-bulge, sometimes also with clusters of stars in the nucleus (Böker et al 2002). Galaxies viewed edge-on and extremely thin galaxies are common (Kautsch et al 2006). According to the statistics of more than 100,000 galaxies from the Sloan survey, more than 20% of bright spiral galaxies have no bulge at all up to a redshift of z=0.03 (Barazza et al 2008). How can these observations be reconciled with the hierarchical scenario?

Classical bulges, which are spheroidal with little flattening, no rotation, and a Sersic profile with index n greater than 3, are known to form in galaxy mergers, both minor and major, which causes them to lose their angular momentum and concentrate their mass. Conversely, pseudo-bulges are the result of the secular evolution of discs, with vertical resonance caused by bars. However, at large redshifts, in the first half of the history of the Universe (z>1), every galaxy experienced a phase where the gas fraction dominated. In these galaxies, where the gas fraction is larger than 50%, the disc is very unstable, and fragments into 3-4 clumps with a baryonic mass of $10^8$-$10^9$ Mo, giving distant galaxies a very piecewise appearance, often described as "clumpy" (cf. Figure 28).

Simulations reveal discs of fragments so massive that their dynamical friction on the dark matter halo slows them down very quickly. In less than a billion years, they spiral toward the center, forming a bulge (Noguchi 1999, Bournaud et al 2007).

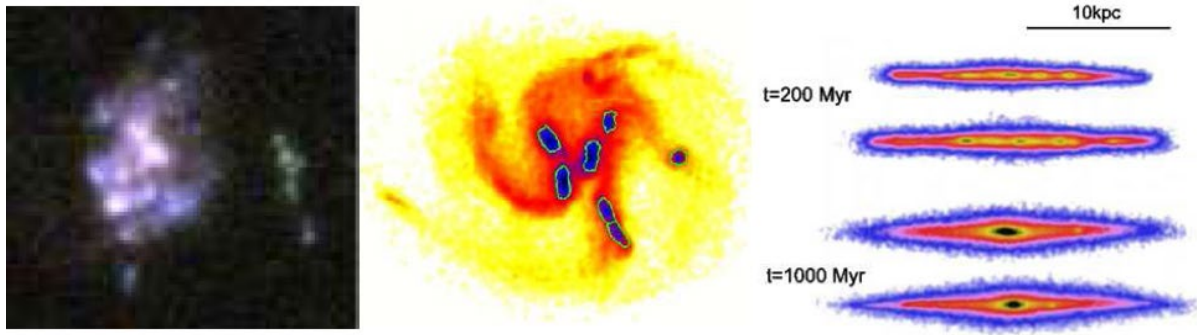

**Figure 28:** Galaxies in the first half of the history of the Universe are dominated by gas, which makes them very unstable and fragmented into "clumpy" objects. On the left, observations of a typical galaxy in the Ultra Deep Field (UDF) of the HST, at z=1.4. In the center and on the right, a simulation, viewed face-on and edge-on, showing the formation of a bulge over a billion years, reproduced from Bournaud et al (2007).

This problem could be resolved by assuming that every spiral galaxy without a bulge or with a classical bulge of very low mass (less than 20% of the total visible mass) has not experienced a major merger since a redshift of z=4 (during around the last 12 billion years). But there is still a classical bulge from the fragmented phase of the first half of the history of the Universe in every galaxy, which does not allow us to account for extremely thin galaxies without any bulge. We would therefore need to invoke a modified gravity model to avoid dynamical friction of the fragments on dark matter halos (Combes 2014). In such a case, the fragments would have time to be destroyed by feedback from star formation and supernovae before being assembled into a bulge.

# 6 Conclusion

Recent large-scale sky surveys have made it possible to study hundreds of thousands of galaxies statistically. This has given rise to a different perspective, one that complements the Hubble sequence classification of galaxies, which remains very useful. Galaxies appear to be separated into two sequences: red and blue, with a relatively unpopulated transition between them, known as the green valley. This bimodality suggests that the transition from the active star formation phase and the inactive red phase occurs rapidly. This rapid transition is closely linked to both the stellar mass of galaxies and their environments.

Various mechanisms have been suggested to explain why star formation is quenched, and numerical simulations have been very helpful. Internal feedback processes originating from either star formation or active galaxy nuclei play a role. But the environment is the most essential factor, as it can permanently suffocate a galaxy by stripping its interstellar gas or suppressing its cold gas supply from cosmic filaments.

One of the driving evolutionary forces, both for mass concentration and star formation, is the development of spiral arms in disc galaxies. There are flocculent spiral galaxies whose arms are irregularly shaped, fragmented, and stochastic; they can be explained by only contagious star formation and differential rotation. But most spirals are coherent, typically with two arms that wind around regularly from the center to the edges. These arms are in fact density waves, generated by the tidal interaction of a companion galaxy, then amplified by swing.

Spirals are sometimes transient, whereas bars are a more robust form of density wave. They allow angular momentum to be exchanged between the inner and outer regions of the galaxy, facilitating the concentration of its mass. They are the driving force of secular evolution, forming rings at the Lindblad resonances, pseudo-bulges by vertical resonance, and spiral structures in the gas.

Spiral galaxies are dominant in the field but give way to lenticular and elliptical galaxies in richer environments such as groups and clusters. These processes likely unfolded gradually, with mergers between galaxies within small groups before these groups then merged within clusters. Many observations are still poorly understood, and there remain dynamical processes to be discovered. Over the next few years, large surveys such as those being conducted with the Euclid satellite, or the LSST on the ground, will further expand our statistical knowledge of galaxies, from one million to ten billion. Giant telescopes both on the ground (both optical with ELTs, Extremely Large Telescopes, and radio with SKA, the Square Kilometer Array) and in space (JWST – James Webb Space Telescope) will allow a fine study of galaxies with high sensitivity and spatial resolution extending up to the confines of our horizon.

## 7   Bibliography of Chapter 4

Athanassoula, E.: 2002, ApJ 569, L83
Baldry, I.K., Glazebrook, K., Brinkmann, J. et al.: 2004, ApJ 600, 681
Baldry, I. K., Balogh, M. L., Bower, R. G. et al.: 2006, MNRAS 373, 469
Barazza, F. D., Jogee, S., Marinova, I.: 2008, ApJ 675, 1194
Baugh, C.M.: 2006, RPPh 69, 3101
Bigiel, F., Leroy, A., Walter, F. et al.: 2008, AJ 136, 2845
Block, D. L., Bournaud, F., Combes, F. et al.: 2002, A&A 394, L35
Böker, T., Laine, S., van der Marel, R. P. et al.: 2002, AJ  123, 1389
Bournaud, F., Combes, F.: 2002 A&A 392, 83
Bournaud, F., Combes, F., Semelin, B: 2005 MNRAS 364, L18
Bournaud, F., Elmegreen, B. G., Elmegreen, D.M.: 20017, ApJ 670, 237
Bureau, M., Freeman, K.C.: 1999 AJ  118, 126
Buta, R., Combes, F.: 1996, Fund. Cosmic Phys. 17, 95
Cappellari, M.: 2016 ARA&A 54, 597
Ciambur, B.C., Graham, A.W.: 2016, MNRAS 459, 1276
Combes, F., Sanders, R.H.: 1981, A&A 96, 164

Combes, F., Debbasch, F., Friedli, D., Pfenniger, D.: 1990, A&A 233, 82
Combes, F.: 2014 A&A 571, A82
Contopoulos, G., Papayannopoulos, T.: 1980, A&A 92, 33
Cowie, L. L., Songaila, A., Hu, E. M., Cohen, J. G.: 1996, AJ 112, 839
Croton, D. J., Springel, V., White, S.D.M. et al: 2006, MNRAS 365, 11
Debattista, V., Sellwood, J.A.: 2000 ApJ 543, 704
Dekel, A, Silk, J: 1986 ApJ 303, 39
Dekel, A, Birnboim, Y.: 2006 MNRAS 368, 2
Di Matteo, P., Gomez, A., Haywood, M. et al.: 2015, A&A 577, A1
Dobbs, C., Baba, J.: 2014, PASA 31, 35
Dominguez Sanchez, H., Huertas-Company, M., Bernardi, M. et al.: 2018, MNRAS 476, 3661
Dressler, A.: 1980, ApJ 236, 351
Driver, S. P., Allen, P. D., Graham, Alister. W. et al.: 2006, MNRAS 368, 414
Eggen, O.J., Lynden-Bell, D., Sandage, A.R.: 1962, ApJ 136, 748
Eke, V.R., Baugh, C.M., Cole, S. et al 2006, MNRAS 370, 1147
Elmegreen, B. G., Elmegreen, D. M.: 1983, ApJ 267, 31
Eskridge, P. B., Frogel, J. A., Pogge, R. W. et al.: 2002, ApJS 143, 73
Fall, S. M., Romanowsky, A. J.: 2013, ApJ, 769, L26
Fisher, D. B., Drory, N.: 2010, ApJ 716, 942
Freeman, K.C.: 1970 ApJ 160, 811
Friedli, D., Martinet, L.: 1993, A&A 277, 27
Garcia-Burillo, S., Sempere, M. J., Combes, F., Neri, R.: 1998, A&A 333, 864
Gerola, H., Seiden, P.E.: 1978 ApJ 223, 129
Helmi, A., White, S.D.M: 1999 MNRAS 307, 495
Hohl, F.: 1971 ApJ 168, 343; 1976, AJ 81, 30
Hopkins, P. F., Quataert, E.: 2010, MNRAS 407, 1529
Jenkins, A., Frenk, C.S., White, S.D.M et al 2001, MNRAS 321, 372
Kalnajs, A.J.: 1973, PAS Au 2, 174
Kauffmann, G., Heckman, T. M., White, S. D. M. et al.: 2003, MNRAS 341, 54
Kautsch, S. J., Grebel, E. K., Barazza, F. D., Gallagher, J. S.: 2006, A&A 445, 765 and 451, 1171
Kormendy, J., Fisher, D.B.: 2008 ASPC 396, 297
Kormendy, J., Drory, N., Bender, R., Cornell, M. E.: 2010, ApJ.723, 54
Lin, C. C., Shu, F. H.: 1964, ApJ 140, 646
Lintott, C. J., Schawinski, K., Slosar, A. et al.: 2008, MNRAS 389, 1179
Madau, P., Dickinson, M.: 2014 ARA&A 52, 415
Martig, M., Bournaud, F., Teyssier, R., Dekel, A.: 2009, ApJ 707, 250
Martinez-Valpuesta, I., Shlosman, I., Heller, C.: 2006, ApJ 637, 214
Milgrom, M.: 1983 ApJ 270, 365
Navarro, J.F., Frenk, C.S, White, S.D.M.: 1997 ApJ 490, 493
Noguchi, M.: 1999 ApJ 514, 77
Ostriker, J.P., Peebles, P.J.E.: 1973, ApJ 186, 467
Peng, Y., Lilly, S. J., Kovac, K. et al: 2010, ApJ 721, 193
Portail, M., Wegg, C., Gerhard, O.: 2015, MNRAS 450, L66
Raha, N., Sellwood, J. A., James, R. A., Kahn, F. D.: 1991, Nature 352, 411
Renaud, F., Bournaud, F., Emsellem, E. et al.: 2013, MNRAS 436, 1836

Schawinski, K., Urry, C.M., Simmons, B. D. et al.: 2014, MNRAS 440, 889
Searle, L., Zinn, R.: 1978 ApJ 225, 357
Sellwood, J.A.: 2014 RvMP 86, 1
Shlosman, I., Frank, J., Begelman, M. C.: 1989, Nature 338, 45
Sweet, S. M., Fisher, D., Glazebrook, K. et al.: 2018, ApJ, 860, 37
Tiret, O., Combes, F.: 2007 A&A 464, 517
Tiret, O., Combes, F.: 2008 A&A 483, 719
Toomre, A.: 1981, Proceedings of the Advanced Study Institute, Cambridge University Press, p. 111-136.
Toomre, A, Toomre, J.: 1972, ApJ 178, 623
Weinzirl, T., Jogee, S., Khochfar, S. et al.: 2009, ApJ 696, 411
Wuyts, S., Förster Schreiber, N. M., van der Wel, A., et al.: 2011, ApJ 742, 96